\documentclass[pdftex,12pt]{JHEP3}
\usepackage{amsmath,amsfonts,amsbsy,latexsym,amssymb,amscd,amstext}
\usepackage[pdftex]{epsfig}
\pdfoutput=1

\def\be{\begin{equation}}
\def\ee{\end{equation}}
\def\bea{\begin{eqnarray}}
\def\eea{\end{eqnarray}}
\def\pd{\partial}
\def\a{\alpha}
\def\b{\beta}
\def\g{\gamma}
\def\d{\delta}
\def\m{\mu}
\def\n{\nu}
\def\t{\tau}

\def\l{\lambda}

\def\r{\rho}

\def\e{\epsilon}

\def\vk{\vec{k}}
\def\vp{\vec{p}}
\def\vx{\vec{x}}
\def\vy{\vec{y}}

\def\bi{\begin{itemize}}
\def\ei{\end{itemize}}

\date{May 25th, 2008} \preprint{IFT-UAM/CSIC-10-86\\FTUAM-10-32}
\title{Finite Time Vacuum Survival Amplitude and Vacuum Energy Decay.} \author{Enrique \'Alvarez and Roberto Vidal \\  Instituto de F\'{\i}sica Te\'orica
UAM/CSIC and Departamento de F\'{\i}sica Te\'orica \\ Universidad
Aut\'onoma de Madrid, E-28049--Madrid, Spain \\ E-mail: \email{enrique.alvarez@uam.es }}
\abstract{ The problem  of the vacuum energy decay is studied for both signs of the cosmological constant, through the analysis of the vacuum survival amplitude, defined in terms of the {\em conformal time}, $z$, by ${\cal A}\left(z,z^\prime\right)\equiv \langle\text{vac}\,z|\text{vac}\,z^\prime\rangle$. Transition amplitudes are computed for finite time-span, $Z\equiv z^\prime-z$, and their {\em late time}  behavior (directly related to the putative decay width of the state)  as well as the transients are discussed up to first order in the coupling constant, $\l$.

}
\begin{document}
\newpage
\section{Introduction}
It has been    claimed \cite{Polyakov} that the free energy
corresponding to an interacting theory in de Sitter space
has got an imaginary part that can be  interpreted as some sort of instability.
\par
The way this imaginary part has to be (perhaps naively) computed is by doing the path integral at imaginary time.
In Poincar\'e coordinates of de Sitter space ($dS_n$) in which the metric reads
\[
ds^2={l^2 dz^2-\sum_i (dx^i)^2\over z^2}
\]

this leads to the metric of {\em euclidean (anti) de Sitter space} ($EAdS_n$)  with metric
\[
ds^2={l^2 dz^2+\sum_i (dx^i)^2\over z^2}
\]
(and not to the metric on the sphere $S_n$), and the corresponding free energy has been computed by us up to the one loop order in the first paper of \cite{Alvarez}, where no imaginary part was find to that order.

Some general arguments can be advanced, however, supporting that a nonvanishing result should be found to higher loop order.
Namely, 
in the second paper of  \cite{Alvarez} it was pointed out that (if it were applicable) the optical theorem relates the (technically quite difficult) computation
of the imaginary part of the free energy to a much simpler tree level calculation, id est, the vacuum decay into identical particles \cite{Higuchi}. A related phenomenon is the decay of a particle into several identical particles \cite{Bros}. Besides, there is no reason for this effect to be restricted to de Sitter space; on the contrary, it would be natural to expect it to be quite generic.
\par

It is true, however, that all our intuition is based upon flat space examples, with the ensuing asymptotic regions, and S-matrix elements that can be computed through LSZ techniques. Outside this framework it is not even known how to define a particle to be decayed into nor the interacting vacuum $|\text{vac}\rangle$ in the absence of a well-defined energetic argument.

A related issue is the study of the time dependence of transition amplitudes. The linear dependence in time is one of the key aspects of Fermi's golden rule. The fact that is problematic in curved space, where there is no naturally preferred coordinate system in general, has been remarked in \cite{Bros}. It is to be stressed that use of non-cartesian coordinates is not without problems even in flat space-time, and this is even more true about polar coordinates in field space. One of the purposes of the present work is to examine this problem, by computing overlaps between states that differ by a finite time in whatever coordinate system we are using. 

The paper is organized as follows. In section \ref{review} we review some results regarding transition amplitudes in both flat and curved space-time, and we present the basic quantum-mechanical formalism that we will use in our calculations. It is based in the functional Schr\"odinger picture for finite time intervals. In section \ref{flatspace}, we put our techniques to work  in order to recover the standard quantum  evolution in Minkowski space-time. This we do in order to check our formalism, and to compare with ordinary quantum  mechanics (as opposed to field theory). In section \ref{desitter} we apply this formalism to de Sitter (and briefly to anti de Sitter) to examine the (conformal time)  dependence of its transition amplitudes.
Some technical details have been relegated to the appendixes
\newpage
\section{Overlaps and evolution}\label{review}
 In flat n-dimensional space-time \cite{Weinberg} (where energy conservation holds) differential transition amplitudes
\footnote{ It is not without interest to recall how the linear dependence in time appears on quantum mechanical survival amplitudes  using old fashioned time-dependent perturbation theory (confer \cite{Sakurai}).
We start with eigenstates of a {\em free} hamiltonian
\[
H_0 |k\rangle=E_k^0 |k\rangle
\]
and expand the full wavefunction as
\[
\psi(t,\vx)\equiv\sum_k c_k(t)u_k(\vx)e^{-i E_k^0 t}.
\]
Schr\"odinger's equation
\[
i{\pd\over \pd t}|\psi\rangle=(H_0+H_1)|\psi\rangle
\]
then demands that
\[
\sum_k H_1 c_k u_k e^{-i E_k^0 t}= i \sum_{ k^\prime}\dot{c}_{k^\prime} u_{k^\prime} e^{-i E_{k^\prime}^0 t}
\]
which implies
\[
\dot{c}_k=-i\sum_{k^\prime} c_{k^\prime}\langle k|H_1 (t)|k^\prime\rangle e^{i(E_k-E_{k^\prime}) t}
\]
When the initial state is an energy eigenstate of the unperturbed hamiltonian, id est, such that $c_k(0)=\d_{k p}$, then,
\[
c_k(T)=-i\int_0^T dt^\prime\langle k|H^1 (t^\prime)|p\rangle e^{i \omega_{k p} t^\prime}
\] 
where $\omega_{k k^\prime}\equiv E_k-E_{k^\prime}$. Assuming a constant perturbation leads to
\[
c_k(T)={2\over T}~e^{i{\omega_{kp}T\over 2}}~H^1_{kp}~{\text{sin}{\omega_{kp}T\over 2}\over \omega_{kp}}
\]
and the decay probability to an state $|k\rangle$ is
\[
P_k(T)\equiv |c_k(T)|^2=4~|H^1_{kp}|^2~{\text{sin}^2{\omega_{kp}T\over 2}\over \omega_{kp}^2}
\]
In the limit when $T\rightarrow\infty$ this reduces to
\[
P_k(T)\equiv |c_k(T)|^2=4~|H^1_{kp}|^2~T~\d(\omega_{kp})
\]
The survival probability is
\[
P(T)\equiv 1-\int \rho(k)dE_k P_k(T)
\]
where $\r(k)$ is the density of final states.\\
}
from an initial state with $N_i$ initial particles do behave {\em for $T$ large enough} as
\[
d\Gamma\sim T V_{n-1}^{1-N_i}
\]
where $T$ is the time span during which the interaction is turned on, and $V_{n-1}$ is the volume of the codimension one spatial sections of constant time of the system (so that in the particular case of vacuum decay it is proportional to the full n-dimensional volume).

In the opposite limit, Maiani and Testa \cite{Maiani} have shown that there is a divergence at small times $T\rightarrow 0$ which survives even after renormalization. In order to eliminate it, a careful study of the incoming wave packet is necessary. 
They studied in particular the example of an unstable scalar particle of mass $M$ decaying into two other scalar particles with masses $m_1$ and $m_2$.
In the narrow packet approximation for the initial state they were able to prove that
\[
S(t)=2\pi g^2~e^{-i M t}~\left(\a_0+\a_1+\b_1\left(t\right)\right)
\]
 where $M$ is the renormalized mass of the resonant state.
 
 \par
The assumption will be made in section \ref{survivals} that  the quantum mechanical formula
\[
\langle\varphi_f\,t_f|\varphi_i\,t_i\rangle\equiv\langle \varphi_f|e^{-iH(t^\prime-t)}|\varphi_i \rangle\equiv\int_{\varphi_i}^{\varphi_f}\mathcal D\phi\,e^{iS[\phi]}
\]
remains valid in curved backgrounds, where the hamiltonian is generically time dependent.

\par

\subsection{Survival amplitudes.}\label{survivals}
Let us introduce the general formalism first in flat space language, but in such a way that it is easily amenable
to generalizations to curved space.
 The whole aim of the present work is to compute   the overlap between an state $| \text{in}\rangle$  defined at a given time  time $t_i$ and another state $|\text{out}\rangle$ defined at a different time $t_f$ (both times can be finite). This would become S-matrix elements in case $t_i\rightarrow -\infty$, $t_f\rightarrow\infty$ and the interaction (including the one resulting from the background gravitational field, if any) is assumed to be switched off at asymptotic times.
\par
First of all, the survival amplitude is an overlap 
\[
\mathcal{A}\left(t_f,t_i\right)\equiv \langle \psi (t_f)|\psi(t_i)\rangle
\]
If the spate is normalized so that
\[
\langle \psi(t_i)|\psi(t_i)\rangle=1
\]
then the unitary evolution (this is a crucial hypothesis) does preserve the norm, so that
\[
 \langle \psi(t_f)|\psi(t_f)\rangle=1
\]
Then Cauchy-Schwarz's inequality guarantees that
\[
|\mathcal{A}\left(t_f,t_i\right)|\equiv |\langle \psi (t_f)|\psi(t_i)\rangle|\leq |\langle \psi(t_i)|\psi(t_i)\rangle|.| \langle \psi(t_f)|\psi(t_f)\rangle| =1
\]
This means that the quantity ($T\equiv t_f-t_i$)
\[
\Gamma (T)\equiv -{2\over T}\,\log|\mathcal{A}\left(t_f,t_i\right)|
\]
must be positive, and in case it is independent of $T$ in the asymptotic regime, could be rightfully interpreted as the {\em decay width} of the state. We shall refer to it loosely as decay width even when it is not constant.
\par
Survival amplitudes are therefore powerful tools to detect instabilities; they are however somewhat blind to the final state of such decays; we will have no precise information on the decay products. This appears to be an important open problem from this viewpoint.
\par
The first principles path integral formula reads
\be\label{formula}
\left.\mathcal{S}\left(t_f,t_i\right)\right|_J\equiv\left.\langle \text{out}|\text{in} \rangle\right|_J=\int [D\varphi_f][D\varphi_i]~\left. \Psi_{t_f}[\varphi_f]^*\langle \varphi_f\,t_f|\varphi_i \,t_i\rangle\right|_J~ \Psi_{t_i}[\varphi_i]
\ee
where the integration measure $[D\varphi]$ is defined in the space of field configurations at fixed ``time''. The wavefunctionals, which are functionals of this fields, are given by
\bea
&&\Psi_{t_f}[\varphi_f]\equiv\langle \varphi_f\, t_f|\text{out}\rangle\nonumber\\
&&\Psi_{t_i}[\varphi_i]\equiv\langle\varphi_i\,t_i| \text{in}\rangle
\eea
An external source $J$ is introduced as usual in order to treat interactions by functional differentiation.
\par
The problem is then reduced to first computing the wavefunctionals of both states (itself a nontrivial task), and then the {\em field transition amplitude}, which is really the Feynman Kernel, or in modern parlance essentially the {\em Sch\"rodinger functional}
\be\label{kernel}
K[J][\varphi_f\,t_f,\varphi_i\,t_i]\equiv\left.\langle \varphi_f\,t_f|\varphi_i\,t_i\rangle\right|_J
\ee
followed by a final functional integration over the possible values of the fields.

This Schrodinger functional will be computed using the general expression in terms of path integral:
\be\label{def}
K[J][\varphi'\,t',\varphi\,t]=\langle\varphi^\prime\,t^\prime|\varphi\,t \rangle\equiv \int_{ \phi(\cdot,t)=\varphi}^{\phi(\cdot,t^\prime)=\varphi^\prime}{\cal D}\phi \, e^{i \int_{\Sigma_t}^{\Sigma_{t^\prime}} d^n x L(\phi,\pd \phi)}\text{ ,}
\ee
where $\Sigma_t$ is a codimension one hypersurface of constant time.\\

Let us remind some well known facts in quantum mechanics (confer, for example, \cite{Sakurai}). If the initial state $| \text{in},t_i\rangle$ (where we have explicitly indicated the possible time dependence) be it the vacuum state or otherwise, is an eigenstante of the full hamiltonian, then Schr\"odinger's equation imply that the modulus of the {\em survival rate} is equal to one and this is true for any values of $t_i$ and $t_f$:
\[
|{\cal A}_{\text{in}}(t_f,t_i)|\equiv \left|\langle \text{in}\,t_f|\text{in}\,t_i\rangle\right|=1
\]
 It is plain that the survival rate is nothing else that the self-overlap ($| \text{in}\rangle=|\text{out}\rangle$) at finite time interval.\\

This means that the only way an state (vacuum or otherwise) can be unstable is by it being a superposition of energy eigenstates. Then the study of the survival amplitude for finite time is quite useful, because we do not need to know any details of the decay process (which is a complicated thing in the absence of
asymptotically flat regions).
\par
In order for a given state to be unstable it is not enough that the survival rate depends on time (this happens already for a linear superposition of only two energy eigenstates), but that this dependence has to be {\em monotonic} in time. It is enough, for example, that 
\[
\dot{\cal A}\neq 0,~\forall t
\]

The actual dependence of the survival rate in quantum field theory with the time interval is however quite complicated. Besides the divergence at small times uncovered by Maiani and Testa \cite{Maiani} (whose understanding demands  a careful treatment of wavepackets in the initial state), it can be explicitly shown in some models that the behavior is oscillating, except at asymptotic times ($T\equiv t_f-t_i\rightarrow\infty$).

\section{Survival amplitudes in flat space}\label{flatspace}

It has been already advertised that the use of our techniques is best illustrated in the simplest flat space example. It is going to be a rather long computation, so let us now draw its roadmap. There are three steps. The most important one is the computation of Schr\"odinger functional or Feynman Kernel; which is the quantum mechanical transition amplitude between states with well-defined values for the fields. This involves the computation of a determinant with Dirichlet boundary conditions. The final step is to integrate over the boundary values of the fields, weighted by the wavefunctional of the state, which we also need to know at this stage.

\subsection {Wavefunctionals}\label{Wavefunctionals}

This means that the first thing we have to do is to find the wavefunctionals \cite{Jackiw}\cite{Long} of the states $| \text{in}\rangle$ and $|\text{out}\rangle$.
Let us begin with the vacuum.
First of all, in flat space the free field and momentum operators read\footnote{We will work in $n$ dimensions, so the vector notation $\vec x$, $\vec p$, etc. will mean always an integration over $n-1$ variables.}
\bea
&&\phi(\vec x)=\int \frac{d\vec p}{(2\pi)^\frac{n-1}{2}} \frac1{\sqrt{2\omega_p}}\left(a_{\vec p}\,e^{i\vp\vx}+ a_{\vec p}^\dagger \,e^{-i\vp\vx}\right)\nonumber\\
&&\pi(\vec x)=-i\int \frac{d\vec p}{{(2\pi)^\frac{n-1}{2}}} \sqrt{\frac{\omega_p}2}\left(a_{\vec p} \,e^{i\vp\vx}- a_{\vec p}^\dagger \,e^{-i\vp\vx}\right)
\eea
so that
\[
a_{\vec p}={1\over \sqrt{2}}\int \frac{d\vec x}{(2\pi)^\frac{n-1}{2}} e^{-i\vp\vx}\left(\sqrt{\omega_p}\phi(\vec x)+i\frac1{\sqrt{\omega_p}}\pi(\vec x)\right)
\]
in such a way that the vacuum wavefunctional obeys
\[
\int  \frac{d\vec x}{(2\pi)^\frac{n-1}{2}} e^{i\vec p\vec x}\left(\sqrt{\omega_p}\varphi(\vec x)+\sqrt{1\over \omega_p}{\d\over \d\varphi(\vec x)}\right)\langle \varphi|0\rangle=0
\]
so that the solution looks exactly the same as in \cite{Jackiw}\cite{Long}, namely,
\be\label{vacWF}
\Psi_0[\varphi]\equiv \langle \varphi|0\rangle =N\,e^{-{1\over 2}\int d\vec x d\vec y\, \omega(\vec{x},\vec{y})\,\varphi(\vec{x})\,\varphi(\vec{y})}
\ee
with
\[
\omega\left(\vec{x},\vec{y}\right)=\omega\left(\vec{y},\vec{x}\right)=\int {d\vec k\over (2\pi)^{n-1}}e^{i\vk(\vx-\vy)}\omega_k
\]
Using the functional Schr\"odinger's equation
\[
i{\pd \over \pd t}\Psi_0[\varphi,t]={1\over 2}\int d\vec x\left(-{\d^2\over \d\varphi^2}-\eta^{ij}\pd_i\varphi\pd_j\varphi+m^2\varphi^2\right)\Psi_0[\varphi,t]
\]
it is possible to determine 
\[
\Psi[\varphi,t]=N e^{-i E_0 t}\Psi[\varphi]
\]
with the vacuum energy defined as
\[
E_0\equiv {1\over 2}\,\text{tr}\,\omega\equiv {1\over 2}\int d \vec x\, \omega(\vx,\vx)=V_{n-1}\rho_0=\frac{V_{n-1}}2\int {d \vec k\over (2\pi)^{n-1}}\,\omega_k
\]
where the spatial volume is denoted by $V_{n-1}\equiv \int d\vec x $. From the expression above it is plain that $E_0$ is both ultraviolet and infrared divergent.
\par
The time-independent normalization factor is given by
\[
N=\det\left({\omega\over \pi}\right)^{1\over 4}=e^{{V_{n-1}\over 4}\int {d\vec k\over (2\pi)^{n-1}}\log{\omega_k\over \pi}}
\]
It is useful to consider eigenfunctions of the kernel defined such that
\[
\int d\vec y\, \omega(\vx,\vy)f_\e(\vy)\equiv \e f_\e(\vx)
\]

In flat space those are just plane waves
\[
f_\e(\vx)\equiv e^{i\vp\vx}
\]
and the eigenvalue reads
\[
\e\equiv \omega_p
\]
In momentum space the vacuum wavefunctional reads
\[
\Psi_0[\phi]=e^{-{1\over 2}\int \frac{d\vec k}{(2\pi)^{n-1}}~\omega_k \phi_{-k}\phi_k}
\]
where
\[
\phi_k\equiv\int d\vec x~e^{-i\vec{k} \vec{x}}\phi_{\vec{x}}.
\]
Then 
the one-particle state (defined in the non-interacting Fock space)  would be defined as
\bea
&&\psi_1[\phi]\equiv \int d\vec x f_\e(\vx)\langle\phi|1\rangle\equiv \int d\vec x f_\e(\vx)\langle\phi|a^\dagger (\vx)|0 t_1\rangle=\nonumber\\
&&{1\over \sqrt{2}}\int d\vec x d\vec y f_\e(\vx)\left(\sqrt{\omega(\vx,\vy)}-\omega^{-1/2}(\vx,\vy){\d\over \d \phi(\vy)}\right)\Psi_0[\phi]=\nonumber\\
&&\sqrt{2 \e} \int d\vec x f_\e(\vx) \phi(\vx)\Psi_0[\phi]
\eea
so that 
\[
E_1=E_0+\e=E_0+\omega_p
\]

In the general case in which there are no asymptotically flat regions of spacetime the uselfulness of those is quite limited.

\subsection{Inclusion of the interaction}
First we write as usual the interacting kernel in terms of the free one using sources
\[
K[J]=e^{i\int_{t_i}^{t_f}dtd\vec x\,L_I\left({\d\over i\d J}\right)} K_0[J]
\]
where $K_0$ is the kernel that corresponds to the free action
\bea
&S_0[\phi]-J\phi&\equiv \int d^n x \left({1\over 2}(\pd\phi)^2-{m^2\over 2}\phi^2-J(x)\phi(x)\right)=\nonumber\\
&&=\int \frac{d\vec k}{(2\pi)^{n-1}}\int_{t_i}^{t_f} dt\left({1\over 2}|\dot{\phi_k}|^2-{\omega_k^2\over 2}|\phi_k|^2-j_{-k}(t)\phi_k(t)\right)
\eea
In order to perform the functional integration, we follow Sakita \cite{Sakita} and split the field into a {\em classical piece}, $\phi^c_k(t)$ (with boundary conditions yet to be specified) and a {\em quantum part}, $\chi_k(t)$
\[
\phi_k(t)=\phi^c_k(t)+\chi_k(t)
\]
so that we have for the measure ${\cal D}\phi=\mathcal D\chi$, as well as
\be
S_0[\phi]-J\phi=S_0[\phi^c]+S_0[\chi]-J\phi^c+\int \frac{d\vec k}{(2\pi)^{n-1}}\int_{t_i}^{t_f} dt\left(\dot{\phi^c_k}\dot\chi_{-k}-\omega_k^2 \phi^c_k\chi_{-k}-j_k(t)\chi_{-k}(t)\right)
\ee
The last piece can be written as
\bea
&&\int_{t_i}^{t_f} dt\left({d\over dt}\left(\dot{\phi^c_k}\chi_{-k}\right)-\chi_{-k}\ddot{\phi^c_k}-\omega_k^2 \phi^c_k\chi_{-k}-j_k(t)\chi_{-k}(t)\right)=\nonumber\\
&&=\dot{\phi^c_k}(t_f)\chi_{-k}(t_f)-\dot{\phi^c_k}(t_i)\chi_{-k}(t_i)-\int_{t_i}^{t_f} dt~\chi_{-k}\left(\ddot{\phi^c_k}+\omega_k^2 \phi^c_k+j_k(t)\right)
\eea
and choose the {\em classical} field $\phi^c_k$ as the solution of the equation $\ddot{\phi^c_k}+\omega_k^2 \phi^c_k+j_k(t)=0$, so
\be\label{expACT}
S_0[\phi]-J\phi=S_0[\phi^c]-J\phi^c+S_0[\chi]+\int d\vec x \left[\chi(x)\phi^c(x)\right]\Big|_{t_i}^{t_f}
\ee

There is an additional contribution coming from the wavefunctionals in the survival amplitude (\ref{formula}). For Fock vacuum wavefunctionals like (\ref{vacWF}), the exponent, depending on the boundary values of $\phi$, can be written as:
\bea\label{expWF}
&&-\frac12\int \frac{d\vec k}{(2\pi)^{n-1}}\omega_k \left(|\varphi_{fk}|^2+|\varphi_{ik}|^2\right)=\\
&&=-\frac12\int \frac{d\vec k}{(2\pi)^{n-1}} \omega_k \Big(|\phi^c_k(t_i)|^2 +|\chi_k(t_i)|^2+2\chi_{-k}(t_i) \phi^c_k(t_i)+\,(t_f\ \text{term})\Big)\nonumber
\eea

The full monty of boundary terms in the sum of (\ref{expACT}) and (\ref{expWF}) is then
\[
i\dot{\phi^c_k}(t_f)\chi_{-k}(t_f)-i\dot{\phi^c_k}(t_i)\chi_{-k}(t_i)-\omega_k \left(\phi^c_k(t_f) \chi_{-k}(t_f)+\chi_{-k}(t_i) \phi^c_k(t_i)\right)
\]
where the $i$ comes from the one in front of the action, $e^{iS_0}$. This means that if we impose on the classical solution $\phi^c$ the boundary conditions
\begin{align}\label{boundary}
&i\dot{\phi^c_k}(t_f)-\omega_k \phi^c_k(t_f) =0\nonumber\\
&i\dot{\phi^c_k}(t_i)+\omega_k  \phi^c_k(t_i)=0
\end{align}
this boundary terms vanish, and the classical field can be expressed in terms of the {\em finite time Feynman propagator}, $\Delta_T(x,x^\prime)$, so that
\[
\phi^c(x)\equiv -\int d^n x^\prime \Delta_T(x,x^\prime)J(x^\prime)
\]
Taking into account that
\begin{align}
S_0[\phi^c]-J\phi^c&=\int dtd\vec x\frac d{dt}\left(\phi^c\dot\phi^c\right)-\int d^nx \frac12(\square+m^2)\phi^c-J\phi^c=\nonumber\\
&=\frac12\int d\vec x\left(\phi^c\dot\phi^c\right)\Big|_{t_i}^{t_f}-\frac{J\phi^c}2=\\
&=-\int\frac{d\vec k}{(2\pi)^{n-1}}\frac{\omega_k}2(|\phi^c_k(t_f)|^2+|\phi^c_k(t_i)|^2)-\frac{J\phi^c}2\nonumber
\end{align}
(where the boundary conditions obtained in equation (\ref{boundary}) have been used), the first terms cancel precisely with the remaining $|\phi^c|^2$ contribution in (\ref{expWF}). The full classical piece in the exponent is given by an expression quadratic\footnote{The time integration in the definition of $\phi^c$ takes place in the whole real line, while the time integration in the $J\phi^c$ term is constrained within the interval $[t_i,t_f]$. Nevertheless, the result can be proved to be independent of the value of the source $J$ outside this interval.} in $J$:
\[
J\phi^c\equiv\int_{t_i}^{t_f} d^nx J(x)\phi^c(x)=-\int_{t_i}^{t_f} d^nx\int_{t_i}^{t_f} d^nx' J(x)J(x')\Delta_T(x,x')\equiv -J\Delta_T J
\]
and we are left with the following expression for the \emph{free} survival amplitude: 
\be
{\mathcal A}_0(t_f,t_i)|_J=e^{{i\over 2}\int_{t_i}^{t_f}\,d^nx d^nx'\,J(x)\Delta_T(x,x')J(x')}\,{\mathcal A}_0(t_f,t_i)|_{J=0}
\ee

We still have to compute $K_0[0]$ insofar as it depends on the initial and final times, $t_i$ and $t_f$ as well as the boundary conditions $[\varphi_i,\varphi_f]$, and integrate it convoluted with the remaining terms of the wavefunctionals. The exponents of these terms depend only on 
\[
\chi\big|_{t_i}=\varphi_i-\phi^c\big|_{t_i}\ ,\ \chi\big|_{t_f}=\varphi_f-\phi^c\big|_{t_f}
\]
so the integration variables can be shifted:
\be
[D\varphi_i][D\varphi_f]=[D\chi\big|_{t_i}][D\chi\big|_{t_f}]
\ee

\subsection{Classical solutions}

The way the computation has been organized is such that it stems from the careful evaluation of classical solutions with well-defined values for the fields at initial and final times. Let us first examine carefully the situation on flat space, and then build upon that.

Any solution of the free Klein Gordon equation in flat space can be written as
\[\label{flatsol}
\phi(x)=\int {d^4 k\over (2\pi)^4}e^{ikx}g(k)\d(k^2-m^2)=\int {d^3 k\over (2\pi)^4}e^{-i\vec{k}\vec{x}}{1\over\omega}\left(g_+\,cos\,\omega t+i g_- \,sin\,\omega t\right)
\]
where
\[
g_\pm\equiv{g(\omega,\vec{k})\pm g(-\omega,\vec{k})\over 2}
\]
It is not difficult to show that the solution that reduces to $\phi_i(\vec{x})$ at $t=t_i$ and $\phi_f(\vec{x})$ at $t=t_f\equiv t_i + T$ is
given by
\[
\phi_c(x)=\int{d\vec k\over (2\pi)^{n-1}}\,e^{-i\vec{k}\vec{x}}\quad{sin\,\omega_k(t-t_i)\phi_f(\vec{k})-sin\,\omega_k(t-t_f)\phi_i(\vec{k})\over sin\,\omega_k T}
\]

The derivatives at the boundary are fixed and given by

\[
\dot{\phi}_c(t_i,\vec{x})=\int{d\vec k\over (2\pi)^{n-1}}\,e^{-i\vec{k}\vec{x}}\quad \omega_k{\phi_f(\vec{k})-cos\,\omega_k T\phi_i(\vec{k})\over sin\,\omega_k T}
\]
\[
\dot{\phi}_c(t_f,\vec{x})=\int{d\vec k\over (2\pi)^{n-1}}\,e^{-i\vec{k}\vec{x}}\quad \omega_k {cos\,\omega_k T\phi_f(\vec{k})-\phi_i(\vec{k})\over sin\,\omega_k T}
\]

 We will eventually be interested in the limit when $T\rightarrow\infty$. Choosing $t_i=-{T\over 2}$ and $t_f={T\over 2}$ it reads
\bea
&&\phi_c(x)=\int{d^3 k\over 16 \pi^3}\,e^{-i\vec{k}\vec{x}}\quad\left(\left({\sin\,\omega_k t\over \sin{\omega_k T\over 2}}+{\cos\,\omega_k t\over \cos{\omega_k T\over 2}}\right)\phi_f(\vec{k})-\left({\sin\,\omega_k t\over \sin{\omega_k T\over 2}}-{\cos\,\omega_k t\over \cos{\omega_k T\over 2}}\right)\phi_i(\vec{k})\right)=\nonumber\\
&&\int{d^3 k\over 16 \pi^3}\,e^{-i\vec{k}\vec{x}}\quad\left(\left({\sin\,\omega_k t\over \sin{\omega_k T\over 2}}\right)\left(\phi_f(\vec{k})-\phi_i(\vec{k})\right)-\left({\cos\,\omega_k t\over \cos{\omega_k T\over 2}}\right)\left(\phi_f(\vec{k})+\phi_i(\vec{k})\right)\right)
\eea

This formula does not hold when $\phi_i=\phi_f=0$. In this case a necessary condition for a solution to exist is that
\[
t_i-t_f\in {\pi\over \omega}\mathbb{Z}
\]
but this cannot hold true for all  frequencies $\omega_k$. The most we can do is to make the solution vanish at one point.
In this case, the field reads

\[
\phi_c(x)=\int{d^3 k\over 8 \pi^3}\,e^{-i\vec{k}\vec{x}}\, g(k)\,\sin\,\omega_k(t-t_i)
\]

Let us examine the classical action when $g=\l=0$
\bea
&&S_c\equiv\int d^n x\left({1\over 2}\left(\pd_\m\phi_c\right)^2-{m^2\over 2}\phi_c^2\right)=\nonumber\\
&&=\int d^n x {1\over 2}\pd_\m\left(\phi_c\pd^\m\phi_c\right)-{1\over 2}\phi_c\left(\Box\phi_c-m^2\phi_c^2\right)=\int d\vec x {1\over 2}\left.\phi_c\dot\phi_c\right|^f_i=\nonumber\\
&&= \int \frac{d\vec k}{(2\pi)^{n-1}} \frac{\omega_k}{2\sin\omega_k T} \left\{(|(\phi_f)_k|^2+|(\phi_i)_k|^2)\cos\omega_k T-2\text{Re}[(\phi_f)_k(\phi_i)_{-k}]\right\}
\eea

 In the particular case when $T\rightarrow i\infty$ we get

\[
S_c=-i\int d\vec x {1\over 2}\left.\phi_c\pd_0 \phi_c\right|^f_i= -i\int d\vec k \,\omega_k\,\left(\phi_f\left(-k\right)\phi_f\left(k\right)+\phi_i\left(-k\right)\phi_i\left(k\right)\right)
\]

\subsection{Computation of the Schr\"odinger functional.}

Let us compute the Schr\"odinger functional $K_0[0][\chi_i,\chi_f]$ that is, the transition amplitude between states with well-defined values for the quantum fields in the free case. This is the quantity that in norelativistic quantum mechanics is aptly named {\em Feynman's kernel}.
\be
K_0[0][\chi^f\,t_f,\chi^i\,t_i]\equiv \int_{\chi^i}^{\chi^f}{\cal D}\chi\, e^{iS_0[\chi]}
\ee
This functional integral is computed with fixed Dirichlet boundary conditions at the endpoints and its functional form is:
\be
K_0[0][\chi^f\,t_f,\chi^i\,t_i]=e^{iS_c[\chi^i,\chi^f]}\det(\Box+m^2)^{-\frac12}
\ee

This expression can be achieved expanding again around a classical solution, given by:
\be
\chi(t,\vec x)=\int\frac{d\vec k}{(2\pi)^{n-1}}{\chi^i_k\,\sin\omega_k (t_f-t)+\chi_k^f\,\sin\omega_k(t-t_i)\over\sin\omega_k T}e^{-i\vec k\vec x}+\xi(t,\vec x)
\ee
so the classical action read
\bea\label{clasica}
&S_c[\chi^i,\chi^f]&={1\over 2}\int \frac{d\vec k}{(2\pi)^{n-1}}\,\chi_k\dot{\chi}_{-k}\Big|^{t_f}_{t_i}=\\
&&=\int \frac{d\vec k}{(2\pi)^{n-1}}\,{\omega_k\over 2\sin~\omega_k T}\left[\left(|\chi_k^i|^2 +|\chi_k^f|^2\right)\cos\omega_k T-2\text{Re}\, \chi_k^i \chi_{-k}^f\right]\nonumber
\eea
and the only thing that remains is to compute the determinant
\be
\det\left(\Box+m^2\right)\equiv\int_{0}^{0}\mathcal D\xi\,e^{i S_0[\xi]}
\ee
with vanishing boundary conditions $\xi\big|_{t_i}=\xi\big|_{t_f}=0$.

It is quite easy to check that the eigenfunctions are given by
\[
\xi_k=\sqrt{2\over T}\left(\sin~{j\pi \left(t-t_i\right)\over T}\right)e^{i\vec{k}\vec{x}}
\]
with eigenvalues $\l_j=\omega_k^2-{j^2\pi^2\over T^2}$, where $j=1,2,\ldots$ For $k=0$ they are normalized in such a way that
\[
\int_0^T dt \xi_n(t) \xi_m(t)=\d_{nm}
\]
The associated zeta function (confer \cite{AlvarezFaedo} whose coventions we follow here) reads
\[
\zeta(s)\equiv V_{n-1}\int {d\vec k\over (2\pi)^{n-1}} \sum_j \left({k^2+m^2-{j^2\pi^2\over T^2}\over \m^2}\right)^{-s}
\]
where $V_{n-1}\equiv \int d\vec x$.
Now we can use the following identities
\begin{align}
&\int {d\vec k\over (2\pi)^{n-1}} \left(k^2+\a^2\right)^{-s}={\pi^{n-1\over 2}\over (2\pi)^{n-1}}{\Gamma(s-{n-1\over 2})\over \Gamma(s)}\left(\a^2\right)^{-s +{n-1\over 2}}\nonumber\\
&\l^{-s}\Gamma(s)=\int_0^\infty d\t~\t^{s-1}~e^{-\l \t}\nonumber
\end{align}
so that
\begin{align}
\zeta(s)&={\pi^{n-1\over 2}\m^{2 s} V_{n-1}(-1)^{-s +{n-1\over 2}}\over (2\pi)^{n-1}\Gamma(s)} \sum_{j\geq 1}\int_0^\infty d\t\,\t^{s-{n+1\over 2}}\,e^{-\t\left({j^2\pi^2\over T^2}-m^2\right)}=\nonumber\\
&={\pi^{n-1\over 2}\m^{2 s} V_{n-1}(m^2)^{-s +{n-1\over 2}}\over (2\pi)^{n-1}\Gamma(s)} \frac12\int_0^\infty dt\,e^{-t}\,t^{s-\frac{n+1}2}\,\left[\theta_3\left(0,{i t\pi\over m^2 T^2}\right)-1\right]
\end{align}
where we have performed analytic continuation in the mass and $\theta_3(z,\t)$ is the ordinary Jacobi elliptic theta function,
\[
\theta_3\left(z,\t\right)\equiv \sum_{n=-\infty}^\infty~ e^{2\pi i n z}e^{i n^2 \pi\t}
\]

This expression is divergent at $s=0$, precisely the point at which its derivative is wanted. We can remedy using a Poisson resummation in the first summand, id est,
\[
\theta_3\left(0,{it\pi\over m^2 T^2}\right)={mT\over \sqrt{t\pi}}\theta_3\left(0,{im^2 T^2\over t \pi}\right)
\]
and representing by $b^2\equiv -{\pi^2\over m^2 T^2}$,
\[
\zeta(s)\Gamma(s)=\frac{\pi^\frac{n-1}{2}(m^2)^{\frac{n-1}2-s}}{2(2\pi)^{n-1}}\m^{2 s} V_{n-1}\left\{\sqrt{\frac\pi{b^2}}\int_0^\infty dt\,e^{-t}t^{s-\frac{n}2-1}\theta_3\left(0,i\frac{\pi}{b^2 t}\right)-\frac12\Gamma\left(\frac{1-n}2+s\right)\right\}
\]\\

The integral is now convergent for appropiate values of $s$:
\be
\int_0^\infty dt\,e^{-t}t^{s-\frac{n}2-1}\theta_3\left(0,i\frac{\pi}{b^2 t}\right)=2\left(\frac b{|j|}\right)^{\frac n2-s}\sum_{|j|\geq1}K_{\frac n2-s}\left(\frac{2\pi |j|}{b}\right)+\Gamma\left(s-\frac n2\right)
\ee
and then

\begin{align}
\zeta(s)=\frac{\pi^\frac{n-1}{2}(m^2)^{\frac{n-1}2-s}}{2(2\pi)^{n-1}\Gamma(s)}\m^{2 s} V_{n-1}\left\{\frac{4(im T)^{s-\frac n2+1}}{\sqrt{\pi}}\right.&\sum_{j\geq 1}j^{s-\frac n2}K_{\frac n2-s}\left(2im Tj\right)+\\
&\left.+\frac{im T}{\sqrt{\pi}}\Gamma\left(s-\frac n2\right)-\Gamma\left(\frac{1-n}2+s\right)\right\}\nonumber
\end{align}
For even values of $n$:
\begin{align}
\zeta'(0)_{\text{even}}=&\frac{\pi^\frac{n-1}{2}m^{n-1}}{2(2\pi)^{n-1}}V_{n-1}\left\{\frac{4(imT)^{1-\frac n2}}{\sqrt{\pi}}\sum_{j\geq1}j^{-\frac n2}K_{\frac n2}(2imT j)-\right.\nonumber\\
&\left.-\Gamma\left(\frac{1-n}2\right)+\frac{(imT)(-1)^\frac{n}{2}}{\sqrt{\pi}(n/2)!}\left[\gamma-\log(m^2/\mu^2)+\psi^{(0)}(1+n/2)\right]
\right\}
\end{align}\\

The sourceless Feynman kernel is precisely
\be
K_0[0][\chi^f,\chi^i]=e^{iS_c} e^{-\frac12\zeta'(0)}
\ee

\subsection{Flat space survival amplitude}

Let us study the vacuum survival amplitude in the free case performing the relevant integrals by brute force, id est, without introducing any sources for the fields
\bea
&&\mathcal A_0\left(t_f,t_i\right)\equiv \langle 0\,t_f|0\,t_i\rangle=\int [D\varphi_f]\,[D\varphi_i]\,\langle 0\,t_f|\varphi_f\,t_f\rangle \langle \varphi_f\,t_f|\varphi_i\, t_i\rangle
\langle \varphi_i\, t_i|0\, t_i\rangle=  \nonumber\\
&&= |N|^2\,\int [D\varphi_f]\,[D\varphi_i]\,e^{-{1\over 2}\int \varphi_f\omega\varphi_f}\,\langle \varphi_f\,t_f|\varphi_i\,t_i\rangle\,e^{-{1\over 2}\int \varphi_i\omega\varphi_i}\nonumber
\eea
where we have introduced the vacuum wavefunctionals of section (\ref{Wavefunctionals}), and the free Feynman kernel has just been shown to be
\[
K_0[0][\varphi_f\,t_f,\varphi_i\,t_i]\equiv \langle \varphi_f\, t_f|\varphi_i\, t_i\rangle=\det\left(\Box+m^2\right)^{-{1\over 2}}~e^{iS_c[\varphi_i,\varphi_f]} 
\]
where the determinant does not depend upon the boundary values for the field variables.\\

This yields
\bea
&&\mathcal{S}_0(t_f,t_i)=|N|^2\det\left(\Box+m^2\right)^{-{1\over 2}}\,\int [D\varphi_f][D\varphi_i]\cdot \nonumber\\
&& \cdot\exp{\left[{i\int {d\vec k\over (2\pi)^{n-1}}\,\omega_k\,\begin{pmatrix}\varphi_{fk}^*&\varphi_{ik}^*\end{pmatrix}\begin{pmatrix}{i\over 2}+{1\over 2}\cot\,\omega_k T&-{1\over 2\sin\,\omega_k T}\\-{1\over 2\sin\,\omega_k T}&{i\over 2}+{1\over 2}\cot\,\omega_k T\end{pmatrix}\begin{pmatrix}\varphi_{fk}\\\varphi_{ik}\end{pmatrix}}\right]}\nonumber
\eea
We shall dub the functional determinant of the operator
\[
B\equiv\begin{pmatrix}{i\over 2}+{1\over 2}\cot\,\omega_k T&-{1\over 2\sin\,\omega_k T}\\-{1\over 2\sin\,\omega_k T}&{i\over 2}+{1\over 2}\cot\,\omega_k T\end{pmatrix}
\]
the {\em boundary determinant}.

The eigenvalues of this matrix are:
\be
\lambda=-1-i\tan(\omega_kT/2)\ ,\ -1+i\cot(\omega_kT/2)
\ee
so the zeta function we have to consider in irder to compute the boundary determinant is (recovering the $\omega_k$ factor we dropped in these eigenvalues):
\bea
\zeta(s)&=&V_{n-1}\m^{s}\int\frac{d\vec k}{(2\pi)^{n-1}} \left(\frac{\omega_k}2\right)^{-s}\left\{\left(-\frac{e^{i\omega_k T/2}}{\cos\omega_k T/2}\right)^{-s}+\left(i\frac{e^{i\omega_k T/2}}{\sin\omega_k T/2}\right)^{-s}\right\}= \\
&=&\frac{V_{n-1}\m^{s}\Omega_{n-2}m^{n-1-s}}{2^{-s}(2\pi)^{n-1}}\int_1^\infty dx (x^2-1)^\frac{n-3}{2}x^{1-s} \left\{\left(-\frac{e^{ixmT/2}}{\cos xmT/2}\right)^{-s}+\left(i\frac{e^{ixmT/2}}{\sin xmT/2}\right)^{-s}\right\}\nonumber
\eea
Now we can perform use the following expansion:
\be
\frac1{(1+y)^{-s}}=\sum_{j=0}^\infty\frac{\pi_j(s)}{j!}y^j
\ee
where $\pi_0(s)=1$ and $\pi_{j+1}(s)=s\cdot\ldots\cdot(s-j)$. The zeta function is then:
\begin{align}
\zeta(s)&=\frac{V_{n-1}\m^{s}\Omega_{n-2}m^{n-1-s}}{2^{-s}(2\pi)^{n-1}}\int dx (x^2-1)^\frac{n-3}{2}x^{1-s} \left\{\left(-\frac{2}{1+e^{-imTx}}\right)^{-s}+\left(-\frac{2}{1-e^{-imTx}}\right)^{-s}\right\}=\nonumber\\
&=\frac{V_{n-1}\m^{s}\Omega_{n-2}m^{n-1-s}(-1)^{-s}}{(2\pi)^{n-1}}2\sum_{j=0}^\infty\frac{\pi_{2j}(s)}{2j!}I_{2j}(s)
\end{align}
with $I_{2j}(s)=\int_1^\infty dx (x^2-1)^\frac{n-3}{2}x^{1-s}e^{-2imTxj}$.\\

Since we know also that $\pi'_0(0)=1$ and $\pi'_{j+1}(0)=(-1)^j j!$, we have:
\bea
&&\zeta'(0)=\frac1{(2\pi)^{n-1}}\Omega_{n-2}m^{n-1}2\left\{I_0'(0)-\log(- m/\mu)\,I_0(0)-\sum_{j\geq1}\frac1{2j} I_{2j}(0)
\right\}
\eea
The integrals can be calculated then:
\begin{align}
&I_0(s)=\int_1^\infty (x^2-1)^{\frac n2-\frac32} x^{1-s} dx=\Gamma\left(\frac{n-1}2\right)\frac{\Gamma\left(\frac{s-n+1}2\right)}{2\Gamma(s/2)}\nonumber\\
&I_{2j}(0)=\frac1{\sqrt{\pi}}\Gamma\left(\frac{n-1}2\right)(ijmT)^{1-\frac n2} K_{\frac n2}(2ijmT)\nonumber
\end{align}
In the even $n$ case the zeta function corresponding to the boundary determinant is
\be
\zeta^\prime (0)_\text{even}={2 \Omega_{n-2}m^{n-1}\over (2\pi)^{n-1}}\Gamma\left({n-1\over 2}\right)\left\{{1\over 4}\Gamma\left({1-n\over 2}\right)-{\left(imT\right)^{1- n/2}\over 2\sqrt{\pi}}\sum_{j\geq 1} j^{-{n\over 2}} K_{n\over 2}\left(2 i m T j\right)\right\}
\ee

Collecting the results of this paragraph with the ones of the previous one, the vacuum survival amplitude in the even $n$ case reads
\begin{align}
{\cal A}_0\left(t_f,t_i\right)\equiv
\textrm{exp}~\bigg[&-{1\over 2}\frac{\pi^\frac{n-1}{2}m^{n-1}}{2(2\pi)^{n-1}}V_{n-1}\left\{\Gamma\left(\frac{1-n}2\right)+\right.\nonumber\\
&\left.+\frac{(imT)(-1)^\frac{n}{2}}{\sqrt{\pi}(n/2)!}\left[\gamma-\log(m^2/2\mu^2)+\psi^{(0)}(1+n/2)\right]\right\}\bigg]
\end{align}

In is remarkable that the end product of this computation is of the form

\[
{\cal A}_0(t_f,t_i)\propto e^{-i E_0 T}
\]
with the {\em vacuum energy} given by
\bea\label{vacuum}
&&{E_0\over V_{n-1}}\equiv {1\over 2}\frac{\pi^\frac{n-1}{2}m^{n}}{2(2\pi)^{n-1}}\frac{(-1)^\frac{n}{2}}{\sqrt{\pi}(n/2)!}\left[\gamma-\log(m^2/2\mu^2)+\psi^{(0)}(1+n/2)\right]
\eea

The inclusion of the interaction in these considerations can be achieved through Feynman diagrams with finite time propagators built in them. 
\subsection{Adding particles}
If we try to perform the same calculation above for an excited state (id est, an state containing particles), the only change is the boundary wavefunctions in the second determinant:
\be
\int [D\varphi_i][D\varphi_f] \varphi_i(\vec p)^*\varphi_f(\vec p)\exp\left[
\int\frac{d\vec k}{(2\pi)^{n-1}}\,\frac{\omega_k}2(\varphi_i\ \varphi_f)_{-\vec k}\left(\begin{array}{cc}i\cot\omega_k T-1 & -i\csc\omega_k T\\
-i\csc\omega_k T & i\cot\omega_k T-1 
\end{array}\right)\left(\begin{array}{c}\varphi_i\\\varphi_f\end{array}\right)_{\vec k}
\right]
\ee
for the case of an state with a particle of momenta $\vec p$.\\

If we introduce sources coupled to the boundary values of $\phi$, the calculation only needs the addition of a term coming from the derivatives:
\begin{align}
&\int [D\varphi_i][D\varphi_f]\exp\left[
\int\frac{d\vec k}{(2\pi)^{n-1}}\,(\varphi_i\ \varphi_f)_{-\vec k}M_{\vec k}\left(\begin{array}{c}\varphi_i\\\varphi_f\end{array}\right)_{\vec k}+(J^i,J^f)_{-\vec k}\left(\begin{array}{c}\varphi_i\\\varphi_f\end{array}\right)_{\vec k}
\right]=\\
=&\exp\left[-\frac14\int\frac{d\vec k}{(2\pi)^{n-1}}(J^i,J^f)_{-\vec k}M_{\vec k}^{-1}\left(\begin{array}{c}J^i\\J^f\end{array}\right)_{\vec k}\right]\int[D\varphi_i][D\varphi_f]\exp\left[
\int\frac{d\vec k}{(2\pi)^{n-1}}\,(\varphi_i\ \varphi_f)_{-\vec k}M_{\vec k}\left(\begin{array}{c}\varphi_i\\\varphi_f\end{array}\right)_{\vec k}
\right]\nonumber
\end{align}
We can get then the additional factors due to the presence of particles:
\be
M^{-1}_{\vec k}=-\frac i{\omega_k}\left(\begin{array}{cc}1 & e^{-i\omega_k T}\\e^{-i\omega_k T} & 1\end{array}\right)
\ee
\be
\frac\delta{\delta J^i_{-\vec p}}\frac\delta{\delta J^f_{\vec p}}\exp\left[-\frac14\int\frac{d\vec k}{(2\pi)^{n-1}}(J^i_{\vec{k}},J^f_{\vec{k}})M_{\vec k}^{-1}\left(\begin{array}{c}J^i_{\vec k}\\J^f_{\vec k}\end{array}\right)\right]\Bigg|_{J=0}\propto e^{-i\omega_k T}
\ee
and this means that the energy in the exponent $e^{-iE_0T}$ calculated for the vacuum-to-vacuum amplitude increases precisely in $\omega_k$. The reason why we did this check is that the way the linear dependence in time appears is quite different in quantum field theory and in quantum mechanics, as we detail in the next paragraph.

\subsection{Quantum mechanics}
It is useful to contrast the field theoretical calculation above with the quantum mechanical harmonic oscillator \cite{Feynman}

The vacuum survival amplitude for the harmonic oscillator with unit massn is given by
\be
C(T)\equiv \langle 0\,t=T|0\,t=0\rangle=\int\mathcal Dq_i\mathcal Dq_f\,e^{-\omega q_i^2/2}e^{-\omega q_f^2/2}\int_{q_i}^{q_f}\mathcal Dq\,e^{iS[q]}
\ee
Expanding the trajectory around the classical solution, $q=q_c+y$, we have:
\be
C(T)=\int\mathcal Dq_i\mathcal Dq_f\,e^{-\omega q_i^2/2}e^{-\omega q_f^2/2}e^{iS[q_c]}\int_{y(0)=y(T)=0}\mathcal Dy\,e^{iS[y]}
\ee
The action for the classical trajectory can be expressed as:
\be
S[q_c]=\left.\frac 12\,q_c\dot q_c\right|^f_i=\frac{\omega}{2\sin\omega T}\left(q_f(\cos\omega Tq_f-q_i)-q_i(q_f-\cos\omega Tq_i)\right)
\ee
and this is quadratic in the boundary values for $q_c$ so that the boundary integral is gaussian:
\begin{align}
&\int\mathcal Dq_i\mathcal Dq_f\,e^{-\omega q_i^2/2}e^{-\omega q_f^2/2}e^{iS[q_c]}=~\pi\left|\begin{array}{cc}
-\frac\omega2+\frac i2\omega\cot\omega T & -\frac i2\omega\csc\omega T\\ 
-\frac i2\omega\csc\omega T & -\frac\omega2+\frac i2\omega\cot\omega T \end{array}\right|^{-\frac12}=\\
&\pi~\left({1\over 2}\omega^2\left(1-i~\textrm{cot}\omega T\right)\right)^{-1/2}=\pi~\frac{\sqrt{2}}{\omega}\sqrt{i\sin\omega T}e^{-i\omega T/2}\nonumber
\end{align}
The Dirichlet determinant can be computed through discretization, following Feynman's original argument, to be
\[
\textrm{det}\left(-{d^2\over dt^2}+\omega^2\right)^{-1/2}=\left(\frac\omega{2\pi i\sin\omega T}\right)^{1/2}
\]
It is amusing to remark that a zeta function computation gives this same determinant only up to a constant.
Altogether it yields
\[
C(T)=\sqrt{\pi\over \omega}~e^{-{i\over 2}\omega T}
\]
The differences with the quantum field theoretic computation are now clear. The dominant terms for large $T$ come from the Schr\"odinger functional only in the field theoretic case, whereas as a result of cancellations, they come from the boundary determinant in quantum mechanics. We just checked that nevertheless the energies of the excited states are also correctly given in quantum field theory.

\newpage
\section{Survival amplitudes in de Sitter space}\label{desitter}
Let us now turn to the main object of our interest, namely the (in)stability of the vacuum state in (anti) de Sitter space.
We shall mainly use here the de Sitter metric in horospheric (Poincar\'e) coordinates, where $z$ plays the role of {\em conformal time}
\[
ds^2={l^2 dz^2-\d_{ij}dx^i dx^j\over z^2}
\]
The conformal time is positive semidefinite
\[
0\leq z\leq \infty
\]
It is sometimes useful to write $z\equiv e^{-{H t}}$, where the Hubble constant, $H$ is related to the radius by $H\equiv{1\over l}$, so that the metric appears in the {\em steady state} form
\[
ds^2=dt^2-e^{2 H t}\d_{ij}dx^i dx^j
\]
In these coordinates it is plain that in the limit $H\rightarrow 0$ ($l\rightarrow\infty$) flat space is recovered.
\par
We are interested in the survival amplitude of a certain state $| \text{in}\rangle$  between (conformal)  time $z$ and $z^\prime$ (both times can be finite)
\[\label{formula2}
{\cal A}_{\text{in}}\left(z_f,z_i\right)\equiv\langle\text{in}\,z_f|\text{in}\,z_i\rangle=\int [D\varphi_f]\,[D\varphi_i]\,\langle \text{in}\,z_f|\varphi_f\,z_f\rangle\langle \varphi_f\, z_f|\varphi_i\,z_i\rangle \langle\varphi_i\,z_i|\text{in}\,z_i\rangle
\]
\subsection{Wavefunctionals}
We shall expand the free field as
\[
\phi(z,\vec{x})=\int d\vec p\left(a_p v_p(z)e^{i\vec{p}\vec{x}}+ a^\dagger_p v^*_p(z)e^{-i\vec{p}\vec{x}}\right)
\]
and the canonically conjugated momentum
\[
\pi(z,\vec{x})=\int {d\vec p\over l z^{n-2}}\left(a_p v^\prime_p(z)e^{i\vec{p}\vec{x}}+ a^\dagger_p (v^\prime_p)^*(z)e^{-i\vec{p}\vec{x}}\right)
\]

Our modes are normalized by the usual Klein-Gordon invariant scalar product
\[
v'^*_p v_p-v^*_p v^\prime_p=i{z^{n-2}l\over (2\pi)^{n-1}}
\]

The creation and annihilation operators are given by:
\bea
&&a_k=-i \frac{z^{2-n}}{l}\int d\vec x\,e^{-i\vec{k}\vec{x}}\left((v^\prime_k)^*(x)\phi(z,\vec{x})-lz^{n-2}v_k^*(z)\pi(z,\vec{x})\right)\nonumber\\
&&a^\dagger_k=i \frac{z^{2-n}}{l}\int d\vec x\,e^{+i\vec{k}\vec{x}}\left(v^\prime_k(x)\phi(z,\vec{x})-lz^{n-2}v_k(z)\pi(z,\vec{x})\right)
\eea

\par
It is now quite plain (at least formally) how to compute wavefunctions for different states.
Let us begin with the wavefunction of the free (Fock) vacuum. It is defined for appropiate destruction operators and a given conformal time $z$ by
\[
a_k|0\rangle= \int d\vec x  e^{-i\vec{k}\vec{x}}\left(v^\prime_k(z)^*\phi(z,\vec{x})-v_k(z)^*lz^{n-2}\pi(z,\vec x)\right)|0\rangle =0
\]

we are thus led to a differential equation common for the vacuum wavefunction,
\[
\int d\vec x e^{-i\vec{k}\vec{x}}\left(i v_k(z)^*z^{n-2}l{\d\over \d\varphi(\vec{x})}+v^\prime_k(z)^* \varphi(\vec{x})\right)\langle \varphi z |0\rangle=0
\]
The vacuum wavefunctional $\langle \varphi|0\rangle$ is the exact analogue of the Schr\"odinger wavefunction $\psi(q,t)\equiv \langle q|\psi\rangle$, where the completeness relationship $\sum |q\rangle\langle q|=1$ is assumed in a time-independent way. Here we introduce a time-independent basis $|\varphi\rangle$ such that
\[
\int [D\varphi] | \varphi\rangle\langle \varphi|=1
\]
This basis is defined in such a way that the diagonalize the field operator
\[
\hat{\phi}(\bar{z},\vec{x})|\varphi\rangle=\varphi(\vec{x})|\varphi\rangle
\]
at a certain fiducial time, $\bar{z}$. But the basis itself depends on this fiducial time in a nontrivial way, and this we have attempted to represent by writing explicitly the basis as $\langle \phi z|$. It follows that a gaussian ansatz

\[
\langle \varphi z|0 \rangle= N \, e^{-{1\over 2}\int d\vec x d\vec y\, K_z(\vec x,\vec y)\varphi(\vec{x})\varphi(\vec{y})}
\]
is indeed a solution, provided
\[
K_z(\vec{x},\vec{y})={-i\over (2\pi)^{n-1}}\int d\vec p\, e^{i\vec{p}(\vec{x}-\vec{y})}\,\frac1{lz^{n-2}}{v^\prime_p(z)^*\over v_p(z)^*}
\]

This gives a natural definition of non-interacting vacuum state corresponding to the modes $v_p(z)$. The present definition of 
vacuum depends on the modes used, and this in turn depends on the physical setup of the question asked. This is a general problem of quantum field theory in a curved space, not specific to our formalism.\\

Through the functional Schr\"odinger's equation perturbative corrections to the noninteracting vacuum can easily be found. 
The concept of {\em particle} is a delicate one when asymptotically flat regions are absent. A possible  definition of a multiparticle state in the present context is, for example,
\[
\langle \varphi\,z|k_1 \ldots  k_p\rangle\equiv\langle\varphi\,z| a^\dagger_{k_1}\ldots a^\dagger_{k_p}|0\rangle
\]
but it is plain that the usefulness of such a definition is quite limited. 


\subsection{Classical solutions}
 The action for a scalar field in a generic conformally flat space (of which both de Sitter and anti de Sitter are particular instances) can be written in a very simple form. We shall insist for no particular reason in keeping the coordinate $z$ dimensionless, so that the dimensionful coordinates are $x^\m\equiv\left(x^0, x^1,\ldots x^{n-2},l z\right)$. The metric is conformally flat
 
\[
ds^2=a(z)^2~\eta_{\m\n}dx^\m dx^\n
\]
In de Sitter space the coordinate $z$ is timelike, so that it follows that
\[
S_{dS}[\phi]=\int~ l dz~ d\vec x ~ a^n\left({1\over a^2}\left({1\over l^2}(\pd_z\phi)^2-(\vec{\nabla}\phi)^2\right)-{m^2\over 2}\phi^2-{\l\over 4!}\phi^4\right)
\]
In anti de Sitter, owing to the fact that the $z$ coordinate is spacelike, this reads
\[
S_{AdS}[\phi]=\int l dz\,d^{n-1}x ~a^n\left({1\over a^2}\left(-{1\over l^2}(\pd_z\phi)^2+\dot{\phi}^2-(\vec{\nabla}\phi)^2\right)-{m^2\over 2}\phi^2-{\l\over 4!}\phi^4\right)
\]
(where now $\vec{x}$ includes all coordinates except $x^0\equiv t$ and $z$).

We shall actually redefine the quantum field (but keep the same notation for it) in order to shift all depence on the background towards the potential 
\[
\phi_{\text{new}}\equiv a^{n-2\over 2}\phi_{\text{old}}
\]
The lagrangian now reads (remember, now $\phi\equiv \phi_{\text{new}}$)
\[
{\cal L}={1\over 2}(\pd\phi)^2-{m(z)^2\over 2}\phi^2-{g(z)\over 6}\phi^3-{\l(z)\over 24}\phi^4\pm{2-n\over 4 l^2}{d\over dz}\left({\dot{a}\over a}\phi^2\right)
\]
where
\bea
&&m^2(z)\equiv~m^2 a^2\pm\left(1-{n\over 2}\right){\ddot{a}\over a l^2}\mp\left({n\over 2}-2\right)\left({n\over 2}-1\right){\dot{a}^2\over a^2 l^2}\nonumber\\
&&\l(z)\equiv a^{4-n}~\l
\eea
(where $\dot{a}\equiv {d a\over d z}$; the upper signs are for de Sitter space, and the lower ones for anti de Sitter).\\

In both de Sitter and anti de Sitter, $a\equiv {1\over z}$ which gives, paying due attention to the fact that $z$ is dimensionless,
\[
S_{dS}={1\over 2}\int d\vec x\,l dz\left\{{1\over l^2}\,\left(\pd_z\phi\right)^2-(\nabla\phi)^2-{m^2l^2-\frac{n(n-2)}4\over l^2~z^2}\phi^2-{\l\over 12}{z^{n-4}\over l}\phi^4\right\} + \int d\vec x \left. {n-2\over 4 z l} \phi^2\right|_{z_i}^{z_f}
\]
\[
S_{AdS}={1\over 2}\int d^{n-1}x\,l dz\left\{- {1\over l^2}~\left(\pd_z\phi\right)^2+\dot{\phi}^2-(\nabla\phi)^2+{m^2~l^2+\frac{n(n-2)}4\over l^2~z^2}\phi^2-{\l\over 12}{z^{n-4}\over l}\phi^4\right\}
\]
where care has been taken to keep in de Sitter all boundary terms for future use. Incidentally, those are totally irrelevant for anti de Sitter, because we are only integrating the time variable over a finite time interval; but they are quite important for de Sitter space, because they enforce a change in Feynman's propagator as explained in the appendix in some detail.\\

It is amusing to remark that up to a constant factor the $\g$ factor defined in the appendix as $\g\equiv {n-2\over 2 z m l }$ is just the de Sitter temperature $T\equiv{1\over 4\pi l}$
\[
\g~m = {n-2\over  z }2 \pi T
\]
It is well known that this temperature is associated to the unavoidable presence of a horizon because of the lack of a globally timelike Killing vector \cite{GibbonsHawking}. This coincidence is due to the fact that there is a single energy scale in de Sitter space.
\par
This action can (and will) be interpreted as a Minkowskian action for a massive field, with a time-dependent potential given by

\[\label{potential}
V\left(z,\phi\right)\equiv -L_I\left(z,\phi\right)\equiv{m^2l^2\left(1-z^2\right)\mp \frac{n(n-4)}4\over 2l^2z^2}\phi^2+{\l\over 24} z^{n-4}\phi^4
\]
The perturbation is a time dependent one for de Sitter space; whereas space dependent in anti de Sitter.
In this split between free and interacting hamiltonian all information on the curvature of the space has been dumped into the potential term.

\subsubsection{A different split between free and interaction terms.}
It can be more convenient for some purposes to keep all quadratic terms (dimension two operators) in the free lagrangian and treat higher dimensional operators as interacting terms.

The scalar field action in de Sitter space
\[
S=\int_{z_i}^{z_f} {l dz d\vec x\over z^n}\left({z^2\over l^2}\left((\pd_z\phi)^2-l^2(\vec{\nabla}\phi)^2\right)-{m^2\over 2}\phi^2-J \phi\right)
\]
reads in momentum space
\[
\phi(z,\vec{x})=\int {d\vec  k\over (2\pi)^{n-1}} \,\phi_k(z)\,e^{i\vec{k}\vec{x}}
\]

\[
S=\int_{z_i}^{z_f} {l dz d\vec k\over z^n}\left({z^2\over l^2}\left(|\pd_z\phi_k|^2+l^2 k^2|\phi_k|^2\right)-{m^2\over 2}|\phi_k|^2-J_{-k} \phi_k\right)
\]
Writing as in flat space,
\[
\phi_k=\phi^c_k+\chi_k
\]
\bea
&&S=\int_{z_i}^{z_f} {l dz d\vec k\over z^n}\left({z^2\over l^2}\left(|\pd_z\phi^c_k|^2+l^2 k^2|\phi^c_k|^2\right)-{m^2\over 2}|\phi^c_k|^2-J_{-k}\left( \phi^c_k+\chi_k\right)+\right.\nonumber\\
&&\left. {z^2\over l^2}\left(|\pd_z\chi_k|^2+l^2 k^2|\chi_k|^2\right)-{m^2\over 2}|\chi_k|^2+{2 z^2\over l^2}\left( \pd_z\phi^c_k \pd_z\chi_{-k}+ l^2 k^2 \phi^c_k \chi_{-k}\right) -m^2\phi^c_k \chi_{-k} \nonumber                                                                                     \right)
\eea

the equation of motion reads 
\[
\phi^{\prime\prime}_k-{n-2\over z}\phi_k^\prime+{l^2\left(k^2+m^2\right)\over z^2}\phi_k=0
\]
whose general solution reads
\[
\phi_k(z)= z^{n-1\over 2}\left(C_1 J_{\left({n-1\over 2}\right)^2-m^2 l^2}\left(klz\right)+C_2 Y_{\left({n-1\over 2}\right)^2-m^2 l^2}\left(klz\right)\right)
\]

To find the solution that reduces to $\phi_1(\vec{x})$ at $z=z_1$ and to $\phi_2(\vec{x})$ at $z=z_2$, let us define
\bea
&&J_{1,2}\equiv J_{\left({n-1\over 2}\right)^2-m^2 l^2}\left(kl z_{1,2}\right)\nonumber\\
&&Y_{1,2}\equiv Y_{\left({n-1\over 2}\right)^2-m^2 l^2}\left(kl z_{1,2}\right)
\eea
Then
\bea
&&C_1={Y_2 \phi_1(k) z_1^{1-n\over 2}-Y_1\phi_2(k)z_2^{1-n\over 2}\over J_1 Y_2-J_2 Y_1}\nonumber\\
&&C_2={-J_2 \phi_1(k) z_1^{1-n\over 2}+J_1\phi_2(k) z_2^{1-n\over 2}\over J_1 Y_2-J_2 Y_1}
\eea
and
\[
\phi_k^\textrm{clas}(z)= C_1 J_\n\left(k l z\right)+C_2 Y_\n\left(k l z\right)
\]
where $\n\equiv \left({n-1\over 2}\right)^2-m^2 l^2$.
For any classical solution the action on shell is given by
\[
S_c=\int d\vec  x {z^{2-n}\over l^2}\phi\pd_z\phi|^2_1=(2\pi)^{n-1}\int{d\vec  k\over l^2}\left.{1\over z^{n-2}}\phi_k(z)(\pd_z \phi_k)(z)\right|_1^2
\]

Let us now find the solution of the equation of motion with a delta-function source and Dirichlet boundary conditions.

Using the fact that the Wronskian
\[
W[J_\n(z),Y_\n(z)]=2~{e^{\pi i\n}\over z \textrm{cos}\n \pi}
\]
the Dirichlet propagator is easily found to be
\[
\Delta^D_k(z,z_0)={1\over 2}~e^{-\pi i \n}~k l z_0 \theta\left(z-z_0\right)\textrm{cos}\pi \n~\left(-Y_\n\left(k l z_0\right)~J_\n\left(k l z\right) + J_\n \left(k l z_0\right)~Y_\n\left(k l z\right)\right)+\phi^\textrm{clas}_k\left(z\right)
\]

We shall develop the expansion associated to this propagator in a forthcoming work.

\subsection{The Schr\"odinger functional}
%
 The Schr\"odinger functional is given by finite-time Feynman's diagrams with position-dependent vertices
\[
\mathcal A(z_f,z_i)|_J=e^{-i\int_{z_i}^{z_f} ldzd\vec x\,V\left(z,i{\d\over l\d J}\right) }\,e^{{i\over 2}\int_{z_i}^{z_f}d^nx d^nx'\,J(x)\, \Delta_T(x,x^\prime)\,J(x^\prime)}\Big|_{J=0}\,\mathcal A_0|_{J=0}
\]
To a given order in perturbation theory, it corresponds to vacuum diagrams (the same that contribute to the usual vacuum energy) computed with finite time Feynman propagators. The important thing to notice is that the only dependence on the boundary values of the fields stems from the classical action.\\

\par
 The first diagram to be computed is the ``circle'', which is simply:
\be
M_{0,0}=\int_{z_i}^{z_f}ldz\int d\vec x \frac{d\vec k}{(2\pi)^{n-1}}\Delta_T(k)[z,z]
\ee
It is to be remarked that even this diagram carries some information about the curvature of the space through the $\g$ terms in the propagator.\\

We can take advantage of the specific form of the said propagator, in the sense that only the first coefficient (\ref{cos}) contributes to the simple diagrams we will consider. In this case, the amplitude reads:
\begin{align}
M_{0,0}=V_{n-1}\int\frac{d\vec k}{(2\pi)^{n-1}}\frac{e^{-2 i l \omega_k  Z} \left(m (\gamma_i-\gamma_f)+e^{2 i l \omega_k  Z} \left(4 l \omega_k ^2 Z+m (\gamma_f-\gamma_i) (1-2 i l \omega_k  Z)\right)\right)}{4 \omega_k ^2 (m (\gamma_i-\gamma_f)-2 i \omega_k )}
\end{align}
where we have neglected the product $\gamma_i\gamma_f$.\\

In the limits of large and small $Z$, (physically, the relevant quantity is $Z m l$)  we have:
\begin{align}
&M_{0,0}\stackrel{Z\to\infty}{\longrightarrow}\frac{V_{n-1}\Omega_{n-2}m^{n-3}}{2(2\pi)^{n-1}}\left\{imlZ\,J\left(\frac{n-3}2,0\right)+\frac{\gamma_i}2I_{00}(Z)\right\}\nonumber\\
&M_{0,0}\stackrel{Z\to0}{\longrightarrow}\frac{V_{n-1}\Omega_{n-2}m^{n-2}}{2(2\pi)^{n-1}}\left\{i lZ \,J\left(\frac{n-3}2,0\right)+\frac{\gamma_i}{2z_i}lZ^2\,J\left(\frac{n-3}2,-1\right)\right\}
\end{align}

where
\begin{align}
&J(a,b)=\int_1^\infty dx (x^2-1)^a\,x^b\nonumber\\
&I_{00}(Z)=\int_1^\infty dx(x^2-1)^\frac{n-3}{2}\frac{(e^{-2imlZx}-1)}{(\gamma_i-2ix)x}
\end{align}

The second diagram is the same as before, but with a mass insertion:
\be
M_{1,0}=-\int_{z_i}^{z_f}ldz\int d\vec x \frac{d\vec k}{(2\pi)^{n-1}}\Delta_T(k)[z,z](\alpha+\frac\beta{z^2})
\ee 
with $\alpha=-m^2/2$ and $\beta=(m^2-n(n-2)/4l^2)/2$. We have a part proportional to the first diagram, and a second part proportional to:
\begin{align}
&\int_{z_i}^{z_f} ldz \frac1{2z^2}e^{i l\omega_k (2 z'-z_f-z_i)} \left((2 \omega_k-i \gamma_f m) e^{2 i l \omega_k (z_f-z')}+i \gamma_f m\right) \left(\gamma_i m \left(-1+e^{2 i l \omega_k (z_i-z')}\right)+2 i \omega_k\right)=\nonumber\\
&=\frac{1}{z_f z_i}l \omega_k  e^{-i l \omega_k  (z_f+z_i)} \Big[-2 i l m \omega_k  z_f z_i \left(\gamma_f (\text{Ei}(2 i l z_f \omega_k )-\text{Ei}(2 i l z_i \omega_k ))+\right.\nonumber\\
&\left.+\gamma_i e^{2 i l \omega_k  (z_f+z_i)} (\text{Ei}(-2 i l z_f \omega_k )-\text{Ei}(-2 i l z_i \omega_k ))\right)-m e^{2 i l \omega_k  z_i} (\gamma_f z_f+\gamma_i z_i)+\nonumber\\
&+e^{2 i l \omega_k  z_f} (m (\gamma_f z_f+\gamma_i z_i)+2 i \omega_k  (z_f-z_i))\Big]
\end{align}
where we have neglected again the terms quadratic in the $\gamma$'s.\\

For large $Z$ this contribution is just a constant independent of $Z$, while that for small $Z$ has a linear and a quadratic part. In this last limit, the full amplitude reads:
\be
M_{1,0}\stackrel{Z\to0}{\longrightarrow}-(\alpha+\frac\beta{z_i^2})M_{00}+\frac{V_{n-1}\Omega_{n-2}m^{n-2}}{2(2\pi)^{n-1}z_i^3}\,ilZ^2 J\left(\frac{n-3}2,0\right)
\ee
where the $M_{00}$ has to be understood as the small-$Z$ limit shown above.\\

\par

The third diagram (the first contribution of the self interaction) is given by
\be
M_{1,0}=3i\int_{z_i}^{z_f}ldz\int d\vec x \frac{d\vec kd\vec p}{(2\pi)^{2n-2}}\Delta_T(k)[z,z]\Delta_T(p)[z,z] \lambda(z)
\ee 
with $\lambda(z)=z^{n-4} \lambda/24 $. The diagram is then proportional to:
\begin{align}
&\int_{z_i}^{z_f} ldz \frac{z^{n-4} }{4} \left((2 \omega_k-i \gamma_f m) e^{2 i l \omega_k (z_f-z')}+i \gamma_f m\right) \left((2 \omega_p-i \gamma_f m) e^{2 i l \omega_p (z_f-z')}+i \gamma_f m\right) \cdot\nonumber\\
&\cdot\left(\gamma_i m \left(-1+e^{2 i l \omega_k (z_i-z')}\right)+2 i \omega_k\right) \left(\gamma_i m \left(-1+e^{2 i l \omega_p (z_i-z')}\right)+2 i \omega_p\right) e^{i l (\omega_k+\omega_p) (2 z'-z_f-z_i)}\simeq\nonumber\\
&\simeq -2 i l \omega_k \omega_p e^{-i l (\omega_k+\omega_p) (z_f+z_i)} \Bigg[\gamma_i l^4 m 2^{3-n} \omega_k^4 \omega_p (i l \omega_k)^{-n-1} e^{2 i l (z_f (\omega_k+\omega_p)+\omega_k z_i)} (\Gamma (n-3,2 i l z_f \omega_k)-\nonumber\\
&-\Gamma (n-3,2 i l z_i \omega_k))+\gamma_i l^4 m 2^{3-n} \omega_k \omega_p^4 (i l \omega_p)^{-n-1} e^{2 i l (z_f (\omega_k+\omega_p)+\omega_p z_i)} (\Gamma (n-3,2 i l z_f \omega_p)-\nonumber\\
&-\Gamma (n-3,2 i l z_i \omega_p))-\frac1{n-3}\left(z_f^{n-3}-z_i^{n-3}\right) e^{2 i l z_f (\omega_k+\omega_p)} (m (\gamma_f-\gamma_i) (\omega_k+\omega_p)+2 i \omega_k \omega_p)-\nonumber\\
&-\gamma_f m 2^{3-n} \omega_p (-i l \omega_k)^{3-n} e^{2 i l \omega_p z_f} (\Gamma (n-3,-2 i l z_f \omega_k)-\Gamma (n-3,-2 i l z_i \omega_k))-\nonumber\\
&-\gamma_f m 2^{3-n} \omega_k (-i l \omega_p)^{3-n} e^{2 i l \omega_k z_f} (\Gamma (n-3,-2 i l z_f \omega_p)-\Gamma (n-3,-2 i l z_i \omega_p))\Bigg]
\end{align}

For large $Z$, the leading term is proportional to $Z^{n-3}$:
\begin{align}
M_{0,1}\stackrel{Z\to\infty}{\longrightarrow}\frac{i\lambda V_{n-1}\Omega_{n-2}^2m^{2n-4}}{32(2\pi)^{2n-2}}&\left\{J\left(\frac{n-3}{2}\right)^2\,\frac{lZ^{n-3}}{n-3}+\left[-J\left(\frac{n-3}{2},0\right)^2+\frac{\gamma_i}{ml}I_{01}(Z)\right]lZ^{n-4}\right\}\nonumber\\
M_{0,1}\stackrel{Z\to0}{\longrightarrow}\frac{i\lambda\Omega_{n-2}^2V_{n-1}m^{2n-4}}{32(2\pi)^{2n-2}}\,&lJ\left(\frac{n-3}2,0\right)\Bigg\{-z_i^{n-4}J\left(\frac{n-3}2,0\right)Z+\\
&+\frac{z_i^{n-5}}2\left[i\gamma_iJ\left(\frac{n-3}2,-1\right)-(n-4)J\left(\frac{n-3}2,0\right)\right]\,Z^2\Bigg\}\nonumber
\end{align}
where:
\be
I_{01}(Z)=\int^\infty_1 dx\int^\infty_1dy(x^2-1)^\frac{n-3}{2}(y^2-1)^\frac{n-3}{2}\frac{e^{-2imlZx} y}{x(2xy+\gamma_i(x+y))}
\ee

\par

\subsection{Vacuum Wavefunctionals}\label{44}
When using, as we do, the variables $\phi_{new}$ the appropiate starting point for the vacuum wavefunctional is the Minkowski one,
\[
\Psi_0[\phi]\equiv ~e^{-i~z~\int_u K_{uu}}~e^{-{1\over 2}\int_{x,y}~K_{x,y}\phi_x \phi_y}
\]
The functional Schr\"odinger's equation, which stems from our main hypothesis on the Feynman kernel in curved space reads
\[
i{\pd \Psi[\phi]\over \pd z}=\int_u~\left(-{1\over 2}{\d^2\over \d \phi_u^2}+{m^2(z)\over 2}\phi_u^2+\left(\nabla\phi\right)^2+{\l(z)\over 24}\phi_u^4\right)\Psi[\phi]
\]
It is possible to solve it in a perturbative way in $\l(z)$ and $\Delta\equiv m^2(z)-m^2$ by writing
\[
\Psi[\phi]=\Psi_0[\phi]+\Delta(z)\Psi_{10}[\phi]+\l(z)\Psi_{01}[\phi]
\]
In this way
\bea
&&\psi_{10}[\phi]=z^{2}~e^{i~z\int_u \left(K_{uv}\phi_v\right)^2-K_{uu}-{\omega_k^2\over 2}\phi_k^2}~e^{-{1\over 2}\int_{xy}K_{xy}\phi_x\phi_y}\nonumber\\
&&\left(-i\int_u~{\phi_u^2\over 2}\int_v\left(K_{vw}\phi_w\right)^2\right)~\Gamma\left(-1,iz\int_u\left(K_{uv}\phi_v\right)^2
\right)
\eea

as well as
\bea
&&\psi_{01}[\phi]=z^{-\left(n-4\right)}~e^{iz\int_u \left(K_{uv}\phi_v\right)^2-K_{uu}-{\omega_k^2\over 2}\phi_k^2}~e^{-{1\over 2}\int_{xy}K_{xy}\phi_x\phi_y}\nonumber\\
&&\left(-i\int_u~{\phi_u^2\over 2}\int_v\left(K_{vw}\phi_w\right)^2\right)~(-1)^n\Gamma\left(5-n,iz\int_u\left(K_{uv}\phi_v\right)^2\right)
\eea

It is remarkable that both terms are proportional to $\Psi_0$, so that the total vacuum wavefunctional can be written as
\[
\Psi[\phi]=\Psi_0\left(1+\Delta \d_1\Psi+\l \d_2\Psi\right)
\]
It is also possible to view the functional Schr\"odinger equation as an evolution equation, and assume that at a given conformal time $z=z_0$ the wavefunctional is $\Psi_0[\phi]$, and then compute its future evolution in the conformal time.
This is {\em not} what we have done here.

\subsection{Survival amplitude}

\begin{figure}[h]\label{diagrams1}
\begin{center}
\includegraphics[scale=0.7]{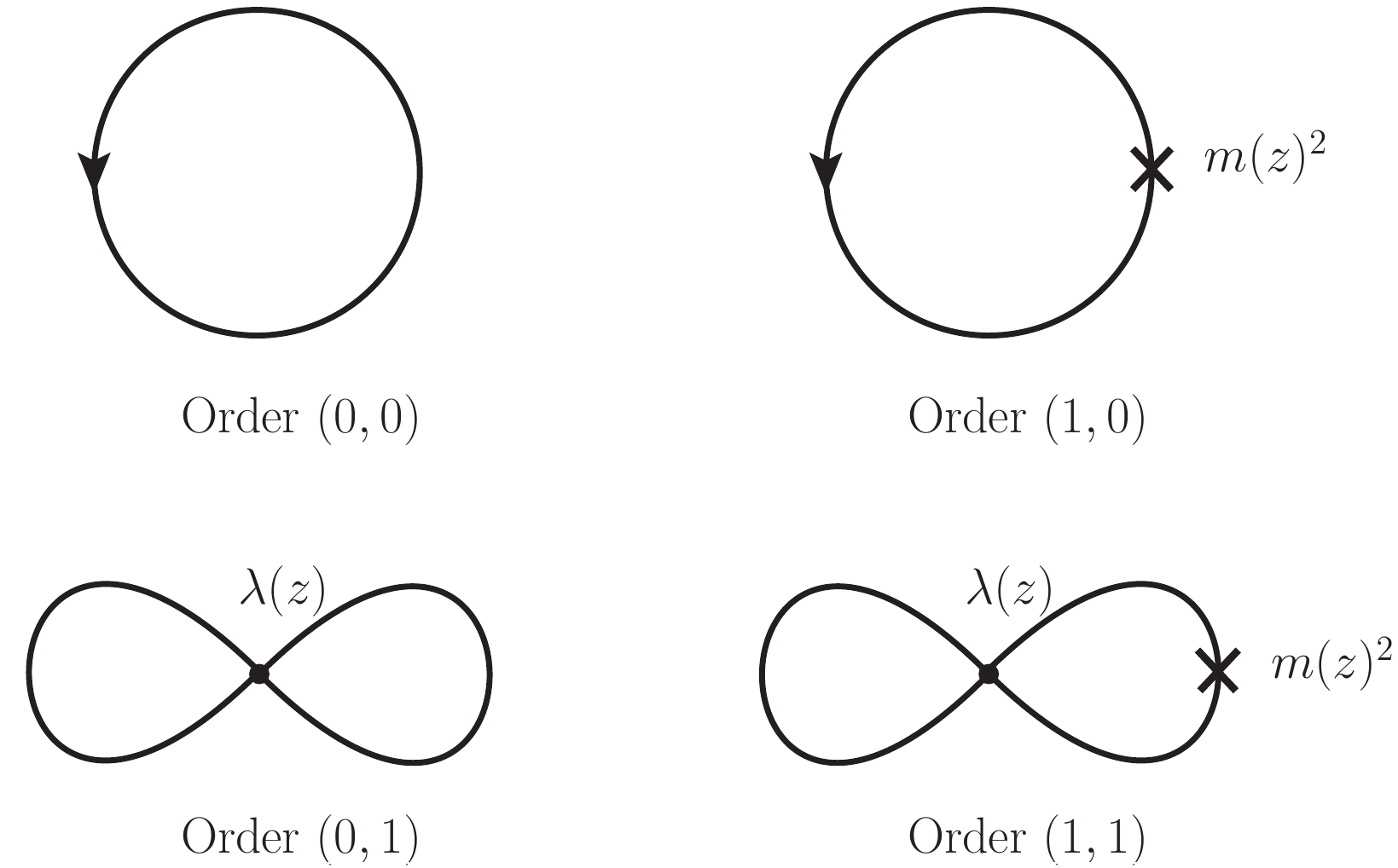}
\caption{The first few diagrams that contribute to the vacuum energy.}
\end{center}
\end{figure}

The only step still left in order to compute the (vacuum) survival amplitude is the integration over the boundary values of the fields, weighed by the vacuum wavefunctions as well as the classical action. We know already from our previous computation (confer equation (\ref{vacuum})) that this contribution is subdominant in the large $Z$ limit, and besides it  preserves the modulus of the (exponentiated) survival amplitude, so that it gives vanishing contribution to the width . 
\par

There are however calculable interaction dependent corrections to the vacuum wavefunction (as to any other wavefunction); they can be obtained through the functional Schr\"odinger's equation to any given order in perturbation theory along the lines of the subsection \ref{44}. We have not attempted to compute the effect of those corrections on the width.

\par
In conclusion,  the value we get for the width of the vacuum state in the asymptotic regime  $Z\rightarrow \infty$ under the approximations of the present work is

\begin{align}
&\Gamma (Z)\stackrel{Z\to\infty}{\longrightarrow}\frac{\alpha V_{n-1}\Omega_{n-2}m^{n-3}\gamma_i}{2(2\pi)^{n-1}Z}\,\text{Re}I_{00}(Z)+\frac{\lambda V_{n-1}\Omega_{n-2}^2m^{2n-5}\gamma_i}{16(2\pi)^{2n-2}}Z^{n-5}\,\text{Im} I_{01}(Z)
\end{align}

\section{Conclusions}
In this paper we have concentrated in computing overlaps between arbitrary states (in particular the vacuum) defined at two different times such that they span a  finite time interval (were this interval infinite they would become $S$-matrix elements, in case  those happen to be well defined). This has been done because there is some initial doubt as to  how to define the good observables (id est, the analogous to the {\em decay rate}, for example \cite{Bros}\cite{Alvarez}) which would presumably involve some sort of square of the overlap matrix elements themselves.
\par
The most important quantity we have analyzed is the survival rate, or self-overlap at  finite (conformal) time span. This in turn determines a {\em decay width} in a straightforward way. It is found that there  some effects already at tree level, which are presumably related to particle creation in the presence of an external non-static gravitational field, but we have not idetintified them unambiguosly. Our computations are consistent with them being transients. They are however of potential physical relevance in the physics of the inflationary epoch. 
\par
At the next order in perturbation theory, there is a new contribution which determines the vacuum width in a precise way in the adequate spacetime dimension (namely, $n=5$ were the dependence of both $I_{01}(Z)$ as well as $I_{00}(Z)$  on their argument subdominant). Further diagrams should be studied before a definite conclusion can be drawn on the main issue.
\par
This computation has been done for a particular wavefunction, which does receive corrections owing to the interaction. Other states can easily be studied within our framework. To the extent that flat space computations are a good guide, we do not expect those improvements to change the physical picture dramatically.
\par
 All the physical quantities studied in this paper turn out to be observer-dependent.
It is not completely clear what could be the physical meaning of some phenomenon which is coordinate dependent (or what amounts to more or less the same thing, observer dependent). There are by now many examples of observer-dependent phenomena even in Minkowski spacetime (of which the Unruh radiation \cite{Unruh} observed by an accelerated oberver in the Minkowski vacuum is perhaps the best known); this does not necessarily mean that their physical meaning is fully understood.

\par

 On the other hand, it is well known that the usual semiclassical approximation to the full quantum theory of the gravitational field interacting with arbitrary matter, namely quantum field theory in a external gravitational field treated classically (upon which the latter identification is based) is only an approximation to the true equations of motion, to wit
 \[
 \langle \text{vac}|{\d S\over \d g^{\m\n}}[g_{\m\n},\psi_i]|\text{vac}\rangle=0
 \]
 Where the total action is the sum of the Einstein-Hilbert part depending on the metric only, the matter part, which depend on the matter fields, denoted here collectively by $\psi_i\quad i=1\ldots N$, and the necessary counterterms, which depend on the metric as well as on the matter fields. Including sources,
 \[
 S[J_{\m\n},J_i]\equiv S_{EH}[g]+S_{\text{matt}}[g,\psi_i]+S_{\text{count}}[g,\psi_i]+i\int d^n x \sqrt{|g|}\left(J^{\m\n} g_{\m\n}+\sum_i J^i \psi_i\right)
 \]
 The equations of motion are always formally true because they can be written in terms of the full partition function
 \[
 Z[J_{\m\n},J_i]\equiv \int {\cal D}g_{\m\n}\prod_i {\cal D}\psi_i~e^{i S[J_{\m\n},J_i]}
 \]
 as
 
 \[
\left.{\d S\over \d g^{\m\n}}\left[{1\over i}{\d\over \d J_{\m\n}(x)},{1\over i}{\d\over \d J_i(x)}\right] Z[J_{\m\n},J_i]\right|_{J=0}=0
 \]
 It remains to give a working definition of the composite operator $g^{\m\n}$, but at the perturbative level this can be done. The state $|\text{vac}\rangle$ is the one obtained through the boundary conditions imposed on the path integral.
 \par
 The semiclassical framework states that this vacuum can be approximated  by the matter vacuum in a fixed gravitational background $\bar{g}_{\m\n}$
 \[
 |\text{vac}\rangle\sim |0_\text{matter}\rangle_{\bar{g}}
 \]

 This can be proven to be the dominant term the first term in a $1/N$ expansion \cite{Hartle} of a theory of gravity interacting with $N$ identical matter species, but it is difficult to believe that this is the {\em only} instance in which this semiclassical approximation is physically reasonable. A general analysis of its validity would be welcome.
 
 \par
 
  Observables in the full quantum gravity theory should presumably be gauge invariant, that is, diffeomorphism invariant, and thus independent on the observer. What seems to be needed here is a gauge invariant definition of vacuum decay.
 \par
  More comprehensive computations are in progress taking into account the dynamics of the gravity sector.

\section*{Acknowledgments}

This work has been partially supported by the European Commission (HPRN-CT-200-00148) as well as by FPA2009-09017 (DGI del MCyT, Spain) and S2009ESP-1473 (CA Madrid). R.V. is supported by a MEC grant, AP2006-01876.  

\newpage
\appendix
\section{Flat space vacuum stability}
Let us first review the reason why those effects vanish in flat Minkowski space, deriving in this way a simple formula for it that subsequently could be  applied to the spaces of our interest in different physical situations.  
To assert that the flat space vacuum state is stable is equivalent to assert that the free energy is formally real in Minkowski space, which through the optical theorem   ensures the stability of flat space versus multiparticle decay. Using LSZ reduction, the S-matrix amplitude for vacuum decay to four identical particles with wave functions $u_{k_1}\ldots u_{k_4}$ reads
\bea
&&\langle k_1 k_2 k_3 k_4|0\rangle=i\int \sqrt{|g(y_1)|} d^n y_1 u_{k_1}^*(y_1)\sqrt{|g(y_2)|} d^n y_2 u_{k_2}^*(y_2)
\sqrt{|g(y_3)|} d^n y_3 u_{k_3}^*(y_3)\nonumber\\
&&\sqrt{|g(y_4)|} d^n y_4 u_{k_4}^*(y_4)
\left(\Box_{y_1}+m^2\right)\left(\Box_{y_2}+m^2\right)\left(\Box_{y_3}+m^2\right)\left(\Box_{y_4}+m^2\right)\times
\nonumber\\
&&\langle 0|T\phi(y_1)\phi(y_2)\phi(y_3)\phi(y_4)|0\rangle\nonumber
\eea
At tree level
\[
\langle 0|T\phi(y_1)\phi(y_2)\phi(y_3)\phi(y_4)|0\rangle=
{\l\over 24}\int d^n y \sqrt{|g(y)|} 
\Delta\left(y_1-y\right)\Delta\left(y_2-y\right)\Delta\left(y_3-y\right)\Delta\left(y_4-y\right)
\]
where the Feynman propagator obeys
\[
\left(\Box+m^2\right) \Delta(x-y)={1\over \sqrt{|g(x)|}} \d(x-y)
\]
In flat space
\[
u_k\equiv{1\over \sqrt{2(2\pi)^{n-1}\omega_k}}e^{-ikx}
\]
so that the net output at tree level is
\[
\int d^n y {1\over\sqrt{\omega_1 \omega_2 \omega_3\omega_4 }} e^{-i\left(k_1+k_2+k_3+k_4\right)y}=(2\pi)^n {1\over\sqrt{\omega_1 \omega_2 \omega_3\omega_4 }}\d\left(k_1+k_2+k_3+k_4\right)
\]

This implies in particular a delta function on the sum of all energies,
\[
\d\left(\sum\,E_i\right)
\]
which does not enjoy support on physical particles. 
\par
 At the same time this gives a simple condition (assuming LSZ reduction is still valid) for this amplitude to be nonvanishing in an arbitrary spacetime, namely, the vacuum is unstable with respect to decay into four particles whenever the fourfold ${\cal O}^{(n=4)}[\phi]$ overlap, where
\[
{\cal O}^{(n)}[\phi]\equiv \int d^n y \sqrt{|g(y)|}u_{k_1}(y) \ldots u_{k_n}(y) 
\]
has got nonvanishing support on physical states. It is plain that this depends on the value of the determinant $g$ (there is always a gauge in which $g=1$) as well as on the set of modes $u_k$. Another quantity of interest in conection to a single particle decay into two or three identical particles is
\[
{\cal O}^{(n-1,1)}[\phi]\equiv \int d^n y \sqrt{|g(y)|}u^*_{k_1}(y) \ldots u_{k_n}(y) 
\]

\section{A first (na\"ive) look at overlaps in de Sitter space.}

Let us examine the overlap with several different {\em particles} (id est,  different coordinate systems) and in different spaces assuming LSZ reduction. The purpose of the present paper was precisely to improve upon this analysis, which we want to briefly present here.
\par
 To  begin with, let us assert that there is no vacuum decay, nor single particle decay into two or three  identical particles in static coordinates (which exist for both de Sitter and anti de Sitter). We call static coordinates ones adapted to the timelike Killing, in such a way the timelike coordinate is ignorable. There may be many different such systems for a given spacetime.
 \par
  The reason is in them the exact modes have got a piece
\[
u_k\sim e^{-i\omega t} f_k
\]
where the functions $f_k$ do not contain the variable time. This is enough to produce a delta function
\[
\d(\sum \omega)
\]
which do not have support on positive energy particles. Using the fact that
\[
\omega^2-m^2\geq 0
\]
it is also possible to show that there is no single particle decay in the static case.
\par
To be specific, de Sitter in static coordinates reads
\[
ds^2=\left(1-{r^2\over l^2}\right)dt^2-{dr^2\over 1-{r^2\over l^2}}-r^2 d\Omega_{n-2}^2
\]
where the radius  of the spacetime is related to  Hubble's constant by
\[
l={1\over H}
\]
and anti se Sitter space in the same coordinates,
\[
ds^2=\left(1+{r^2\over l^2}\right)dt^2-{dr^2\over 1+{r^2\over l^2}}-r^2 d\Omega_{n-2}^2
\]
In anti de Sitter space the horospheric coordinate is spacelike, and it will be denoted by $x$, so that Poincar\'e coordinates admit a FRW-like form
\[
ds^2=-dx^2+e^{2x\over l}\left(dt^2-\sum dy_i^2\right)
\]
which is manifestly static, so that there is no vacuum energy decay here.

De Sitter space in global coordinates (spherical spatial spacelike sections) reads
\[
ds^2=d\t^2-cosh\,\t^2 d\Omega_{n-1}^2
\]

Again, in anti se Sitter in global coordinates, the metric reads
\[
ds^2=cosh^2\,\t d\theta^2-d\t^2-sinh^2\,\t d\Omega_{n-2}^2
\]

Finally, when hyperbolic spacelike sections are considered, de Sitter metric reads
\[
ds^2=d\t^2-sinh^2\,\t\left(d\psi^2+sinh^2\,\psi d\Omega_{n-2}^2\right)
\]

In contrast, anti de Sitter space in the same coordinates yields
\[
ds^2=sinh^2\,\chi d\psi^2-cosh^2\,\chi d\chi^2-cosh^2\,\chi d\Omega_{n-3}^2
\]
which is again explicitly static.

\par
Lest the reader has the impression that anti de Sitter looks atatic in {\em all} coordinates systems, let us mention {\em stereographic coordinates}, in which the metric reads
\[
ds^2=\Omega^2 \eta_{\m\n}dx^\m dx^\n
\]
and for de Sitter space
\[
\Omega\equiv{1\over 1-{x^2\over 4l^2}}
\]
with $x^2\equiv \eta_{\m\n}x^\m x^\n\equiv t^2 -r^2$. For anti de Sitter space
\[
\Omega\equiv{1\over 1+{x^2\over 4l^2}}
\]
In this coordinates, the global staticity of anti de Sitter space is not manifest.

First of all, a general observation \cite{Schrodinger}\cite{Conde}. The link between the field modes and the particle concept is through the WKB approximation. Indeen, the Klein-Gordon equation
\[
\left(g^{\a\b}\nabla_\a\nabla_\b+m^2\right)\phi=0
\]
yields, for
\[
\phi=e^{i{S\over \e}+\ldots}
\]
the {\em mass shell condition}
\[
g^{\a\b}\pd_\a S\pd_\b S=m^2
\]
encoding the definition of positive and negative frequencies for the solutions at hand.
\par
By covariantly deriving the expression above we get
\[
\nabla_\m\pd_\a S g^{\a\b}\pd_\b S=0
\]
Now, for any scalar,
\[
\left(\nabla_\a\nabla_\b-\nabla_\b\nabla_\a\right)S=0
\]
it follows that
\[
g^{\a\b}\pd_\b S\nabla_\a\nabla_\m S=0
\]
namely the geodesic equation. This means that the vector $u^\a\equiv {\pd_\a S\over m}$ is the tangent vector to a geodesic, which in turn implies that the hypersurfaces $S= constant$ are geodesic orthogonal.
\par

\section{Finite time propagators}
The general solution of the Klein Gordon equation
\[
(\Box+m^2)\phi=J
\]
 can be written as
\[
\phi(x)=\int {d\vec k\over (2\pi)^{n-1}}e^{-i\vec{k}\vec{x}}\,\left(a_k\,\cos\,\omega_k\,t+b_k\,\sin\,\omega_k\,t\right)+ \int d^n x^\prime{d^n k\over (2\pi)^n}e^{ik\left(x-x^\prime\right)}\,{P\over -k_0^2+\omega^2_k}J(x^\prime)
\]
The contribution of the principal value is:
\begin{align}
&\int\frac{dk_0}{2\pi}e^{ik_0(t-t')}\frac{P}{-k_0^2+\omega_k^2}=\int\frac{dk_0}{2\pi}e^{ik_0(t-t')}\frac1{2\omega_k}\left(\frac P{k_0+\omega_k}-\frac P{k_0-\omega_k}\right)=\nonumber\\
&=\frac1{2\omega_k}S(t-t')\sin\omega_k(t-t')
\end{align}
where we have used
\[
\int dk\frac P{k-a}e^{ikx}=\int dk \frac{e^{ikx}-e^{ika}}{k-a}=e^{ika}i\int dk\frac{\sin kx}{x}=e^{ika}i\pi S(x)
\]
and $S(x)\equiv\theta(x)-\theta(-x)$ is the sign function. All this leads to
\bea
&\phi(x)=&\int {d\vec k\over (2\pi)^{n-1}}~e^{-i\vec{k}\vec{x}}\,\left(a_k\,\cos\,\omega_k\,t+b_k\,\sin\,\omega_k\,t\right)+\nonumber\\
&&+\int d^n x^\prime\int {d\vec k\over (2\pi)^{n-1}}\,{e^{-i\vec{k}(\vec{x}-\vec{x}^\prime)}\over 2\omega_k}\,S(t-t^\prime)\,\sin\,\omega_k\left(t-t^\prime\right)J(x^\prime)
\eea
In momentum space,
\[
\phi_{-k}(t)=a_k\,\cos\,\omega_k\,t+b_k\,\sin\,\omega_k\,t+{1\over 2\omega_k}
\int d^n x^\prime \,e^{i\vec{k}\vec{x}^\prime}\, S(t-t^\prime)~\,\sin\,\omega_k\left(t-t^\prime\right)J(x^\prime)
\]
\subsection{Dirichlet boundary conditions}

The solution that vanishes at $t=t_i$ as well as at $t=t_f$ then reads
\bea
&&\phi(x)=\int {d\vec k\,d\vec x^\prime\,e^{-i\vec{k}\left(\vec{x}-\vec{x}^\prime\right)}\over (2\pi)^{n-1}2\omega_k\sin\,\omega_k(t_f-t_i)}\int dt^\prime J(t^\prime,\vec{x}^\prime)\Big[
\sin\,\omega_k\left(t_i-t^\prime\right) S\left(t^\prime-t_i\right)\sin\,\omega_k\left(t_f-t\right)+\nonumber\\
&&+\sin\,\omega_k\left(t_f-t^\prime\right) S\left(t^\prime-t_f\right)\sin\,\omega_k\left(t-t_i\right)+ \sin\,\omega_k(t_f-t_i)   \sin\,\omega_k\left(t-t^\prime\right) S\left(t-t^\prime\right)\Big]
\eea
It vanishes for $J=0$, in agreement with previous results.

\par

This means that the correct propagator to be used in the integral over ${\cal D} \xi$ is given by
\bea
&&{\cal D}(x,x^\prime)\equiv\int {d\vec  k\,e^{-i\vec{k}\left(\vec{x}-\vec{x}^\prime\right)}\over 2(2\pi)^{n-1}\omega_k\sin\,\omega_k T}\bigg[
\sin\,\omega_k\left(t_i-t^\prime\right) S\left(t^\prime-t_i\right)\sin\,\omega_k\left(t_f-t\right)+\nonumber\\
&&\sin\,\omega_k\left(t_f-t^\prime\right) S\left(t^\prime-t_f\right)\sin\,\omega_k\left(t-t_i\right)+\nonumber\\
&& \sin\,\omega_k T   \sin\,\omega_k\left(t-t^\prime\right) S\left(t-t^\prime\right)\bigg]
\eea
That is, this is the only solution to the equation
\[
\left(\Box+m^2\right){\cal D}(x,x^\prime)=\d\left(x-x^\prime\right)
\]
such that
\[
\left.{\cal D}(x,x^\prime)\right|_{t=t_i}=\left.{\cal D}(x,x^\prime)\right|_{t^\prime=t_f}=0
\]
\subsection{Feynman boundary conditions}
The boundary conditions for the Feynman propagator are defined by
\bea
&&i\dot{\phi}_k(t_f)=\omega_k \phi_k(t_f)\nonumber\\
&&i\dot{\phi}_k(t_i)=-\omega_k \phi_k(t_i)
\eea
In momentum space
\[
\dot{\phi}_{-k}(t)=-\omega_k a_k \sin~\omega_k t+ b_k \omega_k \cos~\omega_k t+\int_{\mathbb{R}^n} d^n x^\prime~{e^{i\vec{k}\vec{x}^\prime}\over 2\omega_k}S(t-t^\prime)\omega_k\cos~\omega_k (t-t^\prime)J(x^\prime)
\]
(the delta function does not contribute).\\

The boundary conditions are then

\bea
&&-i\omega_k a_k \sin~\omega_k t_f+i b_k \omega_k \cos~\omega_k t_f+i\int_{\mathbb{R}^n} d^n x^\prime~{e^{i\vec{k}\vec{x}^\prime}\over 2\omega_k}S(t_f-t^\prime)\omega_k\cos~\omega_k (t_f-t^\prime)J(x^\prime)
=\nonumber\\
&&\omega_k\left( a_k\,cos\,\omega_k\,t_f+b_k\,sin\,\omega_k\,t_f+
{1\over2 \omega_k}\int_{\mathbb{R}^n} d^4 x^\prime ~e^{i\vec{k}\vec{x}^\prime}\, ~S(t_f-t^\prime)~\,sin\,\omega_k\left(t_f-t^\prime\right)J(x^\prime)\right)\nonumber
\eea
\bea
&&-i\omega_k a_k \sin~\omega_k t_i+ i b_k \omega_k \cos~\omega_k t_i+i \int_{\mathbb{R}^n} d^n x^\prime~{e^{i\vec{k}\vec{x}^\prime}\over 2\omega_k}S(t_i-t^\prime)\omega_k\cos~\omega_k (t_i-t^\prime)J(x^\prime)=\nonumber\\
&&-\omega_k\left(
a_k\,cos\,\omega_k\,t_i+b_k\,sin\,\omega_k\,t_i+{1\over 2\omega_k}
\int_{\mathbb{R}^n} d^4 x^\prime ~e^{i\vec{k}\vec{x}^\prime}\, ~S(t_i-t^\prime)~\,sin\,\omega_k\left(t_i-t^\prime\right)J(x^\prime)\right)\nonumber
\eea\\

This can be written as
\bea
&&\begin{pmatrix}- e^{i \omega_k t_f}&i~e^{i \omega_k t_f}\\ e^{-i \omega_k t_i}&i e^{-i \omega_k t_i}\end{pmatrix}\begin{pmatrix}a_k\\b_k\end{pmatrix}=-{1\over 2\omega_k}\int_{\mathbb{R}^n} d^n x^\prime~e^{i\vec{k}\vec{x}^\prime}\begin{pmatrix}S(t_f-t^\prime)i e^{i \omega_k(t_f-t^\prime)}\\S(t_i-t^\prime)i e^{-i \omega_k(t_i-t^\prime)}\end{pmatrix}J(x^\prime)\nonumber
\eea
and this yields
\bea
&&a_k={i\over 4 \omega_k}~e^{-i\omega_k T}\int d^n x^\prime e^{i\vec{k}\vec{x}^\prime} J(x^\prime)\left(S(t_f-t^\prime)~e^{-i\omega_k(t^\prime-T)}-S(t_i-t^\prime)~e^{i\omega_k(T+t^\prime)}\right)=\nonumber\\
&&{i\over 4 \omega_k}~\int_{\mathbb{R}^n} d^n x^\prime e^{i\vec{k}\vec{x}^\prime} J(x^\prime)\left(S(t_f-t^\prime)~e^{-i\omega_k t^\prime}-S(t_i-t^\prime)~e^{i\omega_k t^\prime}\right)
\eea
\bea
&&b_k=-{1\over 4\omega_k}~\int_{\mathbb{R}^n} d^n x^\prime e^{i\vec{k}\vec{x}^\prime} J(x^\prime)e^{-i\omega_k T}\left(S(t_f-t^\prime)~e^{i\omega_k(T-t^\prime)}+S(t_i-t^\prime)~e^{i\omega_k(T+t^\prime)}\right)=\nonumber\\
&&-{1\over 4\omega_k}~\int_{\mathbb{R}^n} d^n x^\prime e^{i\vec{k}\vec{x}^\prime} J(x^\prime)\left(S(t_f-t^\prime)~e^{-i\omega_k t^\prime}+S(t_i-t^\prime)~e^{i\omega_k t^\prime}\right)
\eea\\

This means that the corresponding classical solution reads
\[
\phi(x)\equiv \int_{\mathbb{R}^n} d^n x^\prime~\Delta_T(x,x^\prime)J(x^\prime)
\]
with $\Delta_T$ the finite time Feynman propagator:
\[
\Delta_T(x,x^\prime)\equiv\int {d\vec k\over(2\pi)^{n-1}}e^{-i\vec{k}(\vec{x}-\vec{x}^\prime)}\Delta_T(k)
\]
\begin{align}
\Delta_T(k)={i\over 4 \omega_k}\Big[& \left(S(t_f-t^\prime)~e^{-i\omega_k t^\prime}-S(t_i-t^\prime)~e^{i\omega_k t^\prime}\right)
\cos~\omega_k t+\nonumber\\
&+i\left(S(t_f-t^\prime)~e^{-i\omega_k t^\prime}+S(t_i-t^\prime)~e^{i\omega_k t^\prime}\right)
\sin~\omega_k t\Big]+\nonumber\\
+{1\over 2\omega_k}&S(t-t^\prime)\sin~\omega_k(t-t^\prime)
\end{align}
It is then plain that in the limit $t_f=-t_i={T\over 2}$ and $T\rightarrow\infty$, 
 Feynman's continuum result is recovered.
\[
\lim_{T\rightarrow\infty} \Delta_T(k)=\Delta_F(k)\equiv {i\over 2\omega_k}\left[\cos~\omega_k (t-t^\prime)-i S(t-t^\prime)\sin~\omega_k (t-t^\prime)\right]
\]

In general it yields
\begin{align}
\Delta_T(k)=\frac i{4\omega_k}\Big([S(t_f-&t')-S(t_i-t')-2]\cos\omega_k(t-t')+\nonumber\\
&+i[S(t_f-t')+S(t_i-t')]\sin\omega_k(t-t')\Big)+\Delta_F(k)
\end{align}

The above results are valid for general sources with arbitrary support. When (as in our case) the support is restricted to the interval $t_i\leq t^\prime\leq t_f$, it is quite easy to check that the finite time propagator with Feynman's boundary conditions coincides exactly with ine usual Feynman's propagator.
\[
\Delta_T(k)\big|_{[t_i,t_f]}=\Delta_F(k)\big|_{[t_i,t_f]}
\]
\subsection{Feynman's propagator including de Sitter boundary terms}
The boundary terns that appear when redefining the field in de Sitter space imply (after the splitting $\phi=\phi_c+\chi$) an addition of
\[
i{n-2\over 2 z_f l}\phi_c (z_f) \chi(z_f)- i{n-2\over 2 z_i l}\phi_c(z_i) \chi(z_i)\equiv i\g_f m \phi_c (z_f) \chi(z_f)- i\g_i m \phi_c(z_i) \chi(z_i)
\]
to the boundary. In order to eliminate those cross-terms the boundary conditions to be imposed are

\bea
&&i\dot{\phi}^c_k(z_f)-\omega_kl\,\phi^c_k(z_f)+i\g_f m l\,\phi^c_k(z_f)=0\nonumber\\
&&i\dot{\phi}^c_k(z_i)+\omega_kl\,\phi^c_k(z_i)+i\g_i m l\,\phi^c_k(z_i)=0
\eea

Let us now can make an slightly different antsatz  for the form of the propagator, namely
\begin{align}
\Delta_T(k)=&\frac1{2\omega_k}S(z-z')\sin\omega_kl(z-z')+\nonumber\\
&+a_k \cos\omega_kl(z-z')+b_k \sin\omega_kl(z-z')
\end{align}
The previous equations for the boundary values give rise to the following coefficients:
\bea
&D_k\equiv&\frac{e^{-i l \omega_k  Z}}{\omega_k   \left(-\gamma_f \gamma_i m^2 \left(e^{2 i l \omega_k  Z}-1\right)-2 i m \omega_k  (\gamma_f-\gamma_i) +4 \omega_k ^2 \right)}\nonumber\\
&a_k=&\frac{D_k}{4}  \left((2 \omega_k-i \gamma_f m) e^{2 i l \omega_k (z_f-z')}+i \gamma_f m\right) \left(\gamma_i m \left(-1+e^{2 i l \omega_k (z_i-z')}\right)+2 i \omega_k\right)\cdot\nonumber\\
&&\cdot(S(z_f-z')-S(z_i-z')) e^{i l\omega_k (2 z'-z_f-z_i)}\nonumber\\
&b_k=&\frac{D_k}4 e^{-i l \omega_k  (2 z'+z_f+z_i)} \Bigg[S(z_f-z') \left(\gamma_f m e^{2 i l z' \omega_k }+e^{2 i l \omega_k  z_f} (\gamma_f (-m)-2 i \omega_k )\right) \Big(e^{2 i l z' \omega_k } (\gamma_i m-2 i \omega_k )+\nonumber\\
&&+\gamma_i m e^{2 i l \omega_k  z_i}\Big)+S(z'-z_i) \left(\gamma_f m e^{2 i l z' \omega_k }+e^{2 i l \omega_k  z_f} (\gamma_f m+2 i \omega_k )\right)\cdot\nonumber\\
&&\cdot\left(e^{2 i l z' \omega_k } (\gamma_i m-2 i \omega_k )-\gamma_i m e^{2 i l \omega_k  z_i}\right)\Bigg]
\eea
where $Z=z_f-z_i$.\\

However, the interaction takes place only in the interval $[z_i,z_f]$, so we should restrict the variables $z$ and $z'$ to be into this interval, so:
\bea\label{cos}
&a_k&=\frac{D_k}{2} e^{i l\omega_k (2 z'-z_f-z_i)} \left((2 \omega_k-i \gamma_f m) e^{2 i l \omega_k (z_f-z')}+i \gamma_f m\right) \left(\gamma_i m \left(-1+e^{2 i l \omega_k (z_i-z')}\right)+2 i \omega_k\right)\nonumber\\
&b_k&=\frac{D_k}{2} m\, e^{-i l \omega_k (2 z'+z_f+z_i)} \left(\gamma_f e^{4 i l z' \omega_k} (\gamma_i m-2 i \omega_k)-\gamma_i (\gamma_f m+2 i \omega_k) e^{2 i l \omega_k (z_f+z_i)}\right)\nonumber\\
\eea

Also, since the exponent of the Feynman Kernel is symmetric in the source $J$, we must symmetrise the propagator.

\newpage

\end{document}